\documentclass[11pt]{article}
\usepackage{slantsc}
\usepackage{lmodern}
\usepackage[pagebackref,colorlinks=true,citecolor=blue]{hyperref}
\usepackage{amsmath} 
\usepackage{amsthm} 
\usepackage{thmtools}
\usepackage{amssymb}	
\usepackage{graphicx} 
\usepackage{multicol} 
\usepackage{multirow}
\usepackage{color}
\usepackage[letterpaper,margin=1in]{geometry}
\usepackage[capitalize]{cleveref}

\usepackage[utf8]{inputenc}
\usepackage[english]{babel}
\usepackage{mathtools}
\usepackage{quantikz}

\usepackage{interval}
\intervalconfig{soft open fences}

\newtheorem{theorem}{Theorem}[section]

\newtheorem{lemma}[theorem]{Lemma}

\newtheorem{proposition}[theorem]{Proposition}
\newtheorem{fact}[theorem]{Fact}

\newtheorem{definition}[theorem]{Definition}

\renewcommand{\braket}[2]{\left< #1 \vphantom{#2} \middle| #2 \vphantom{#1} \right>} 
\newcommand{\ketbra}[2]{\ensuremath{\ket{#1}\!\bra{#2}}}

\DeclarePairedDelimiter\rbra{\lparen}{\rparen}
\DeclarePairedDelimiter\sbra{\lbrack}{\rbrack}
\DeclarePairedDelimiter\cbra{\{}{\}}
\DeclarePairedDelimiter\abs{\lvert}{\rvert}
\DeclarePairedDelimiter\Abs{\lVert}{\rVert}
\DeclarePairedDelimiter\ceil{\lceil}{\rceil}

\let\ket\relax
\DeclarePairedDelimiter\ket{\lvert}{\rangle}
\let\bra\relax
\DeclarePairedDelimiter\bra{\langle}{\rvert}

\newcommand{\tr} {\operatorname{tr}}
\newcommand{\poly} {\operatorname{poly}}
\newcommand{\diag} {\operatorname{diag}}

\newcommand{\rank} {\operatorname{rank}}
\newcommand{\supp} {\operatorname{supp}}
\newcommand{\spanspace} {\operatorname{span}}

\newcommand{\ran} {\operatorname{ran}}

\usepackage{amsmath,amssymb,amsthm,amsfonts,latexsym,mathtools,mathdots,bbm,graphicx,float}
\usepackage{caption,subcaption,ellipsis,xcolor,textcomp,thmtools,thm-restate,mleftright}
\usepackage{cmap}
\crefname{ineq}{inequality}{inequalities}
\creflabelformat{ineq}{#2{\upshape(#1)}#3}
\usepackage{enumitem}

\newcommand{\ignore}[1]{}

\DeclareMathOperator*{\Prob}{\mathbf{Pr}}

\DeclareMathOperator*{\Var}{\mathbf{Var}}

\newcommand{\R}{\mathbb R}
\newcommand{\C}{\mathbb C}

\newcommand{\Z}{\mathbb Z}

\newcommand{\BQP}{\mathsf{BQP}}

\newcommand{\eps}{\varepsilon}

\newcommand{\calA}{\mathcal{A}}

\newcommand{\calC}{\mathcal{C}}
\newcommand{\calD}{\mathcal{D}}
\newcommand{\calE}{\mathcal{E}}

\newcommand{\calH}{\mathcal{H}}

\newcommand{\calO}{\mathcal{O}}

\newcommand{\calS}{\mathcal{S}}

\DeclarePairedDelimiter\norm{\lVert}{\rVert}

\DeclarePairedDelimiterX\diverg[2]{(}{)}{#1 \,\|\, #2}

\newcommand{\dtr}[3]{\mathrm{T}^{#3}\rbra*{#1,#2}}

\newcommand{\power}[3]{\mathrm{F}_{#3}(#1,#2)}
\newcommand{\dpower}[3]{\mathrm{\Phi}_{#3}\diverg{#1}{#2}}

\newcommand{\PureQSD}{\textnormal{\textsc{PureQSD}}\xspace}

\newcommand{\PowerLowQSD}{\textnormal{\textsc{PowerMeanLowRankQSD}}\xspace}

\newcommand{\PowerPureQSD}{\textnormal{\textsc{PowerMeanPureQSD}}\xspace}

\usepackage{algorithm}
\usepackage{algpseudocode}

\usepackage{tabularx}
\usepackage{booktabs}
\usepackage{threeparttable}

\usepackage{adjustbox}

\usepackage{footnotehyper} 
\makesavenoteenv{table}  

\newcommand{\footremember}[2]{%
    \footnote{#2}
    \newcounter{#1}
    \setcounter{#1}{\value{footnote}}%
}

\usepackage{tikz}
\usetikzlibrary{arrows.meta,positioning,fit}

\title{On Estimating Quantum Divergences with $p$-Power Means}

\author{Jinge Bao\footremember{1}{University of Edinburgh. \href{mailto:jinge.bao@ed.ac.uk}{\nolinkurl{jinge.bao@ed.ac.uk}}.} \and
Minbo Gao\footremember{2}{Institute of Software, CAS. \href{mailto:gaomb@ios.ac.cn}{\nolinkurl{gaomb@ios.ac.cn}} or
\href{mailto:gmb17@tsinghua.org.cn}{\nolinkurl{gmb17@tsinghua.org.cn}}.} \and
Qisheng Wang\footremember{3}{Shanghai Jiao Tong University. \href{mailto:QishengWang1994@gmail.com}{\nolinkurl{QishengWang1994@gmail.com}}.}}

\date{}

\begin{document}

\maketitle

\begin{abstract}
    We study the problem of estimating the quantum divergence with $p$-power means, $\Phi_p(\rho\,\|\,\sigma)$. 
    This quantity was investigated by \hyperlink{cite.LL21}{Lam and Le (\textit{Linear Algebra Appl.}\ 2021)}. 
    For any constant $p \in (0, 1)$, we present sample-efficient and query-efficient quantum algorithms for estimating $\Phi_p(\rho\,\|\,\sigma)$ and further show that estimating $\Phi_p(\rho\,\|\,\sigma)$ is $\mathsf{BQP}$-complete in the low-rank case. 

\end{abstract}

\newpage

\tableofcontents

\newpage

\section{Introduction}

Quantum divergences are distance-like measures of the closeness between quantum states. 
Examples include the von Neumann relative entropy \cite{Ume62}, Petz-R\'enyi relative entropy \cite{Pet86,Ren61}, and sandwiched R\'enyi relative entropy \cite{WWY14,MDS+13} (generalizing the Uhlmann fidelity \cite{Uhl76,Joz94}). 
The estimation of quantum divergences usually requires additional assumptions on the quantum states in order to avoid the infinity issue, e.g., \cite{Hay25,LF25}. 
There are also quantum divergences with bounded values, such as the Tsallis relative entropy \cite{Abe03}, for which the estimation was recently studied in \cite{BGW25}.
In addition to the quantum distances such as the trace distance \cite{Hel67,Hel69,Rus94}, Bures distance (induced by the Uhlmann fidelity) \cite{Bur69}, Hellinger distance (induced by the Tsallis relative entropy) \cite{LZ04}, and Hilbert-Schmidt distance \cite{Oza00}, a family of \textit{symmetric} quantum divergences were recently proposed in \cite{LL21}, called quantum divergences with $p$-power means. 
Specifically, in \cite{LL21}, for $p \in \rbra{0, 1}$, the quantum divergence with $p$-power means between two quantum states $\rho$ and $\sigma$ is defined by
\[
    \dpower{\rho}{\sigma}{p} = 1-\power{\rho}{\sigma}{p}=1-\tr\rbra*{\rbra*{\frac{\rho^p+\sigma^p}{2}}^{\frac{1}{p}}},
\]
where $\power{\cdot}{\cdot}{p}$ is the $p$-power mean and they showed that $\dpower{\rho}{\sigma}{p} \in \sbra{0, 1}$ satisfies the data-processing inequality $\dpower{\mathcal{E}\rbra{\rho}}{\mathcal{E}\rbra{\sigma}}{p} \leq \dpower{\rho}{\sigma}{p}$ for any quantum channel $\mathcal{E}$ and $\dpower{\rho}{\sigma}{p}$ has the limiting cases 
\[
    \dpower{\rho}{\sigma}{0} = 1 - \tr\rbra*{e^{\frac{\ln\rbra{\rho}+\ln\rbra{\sigma}}{2}}}, \quad \dpower{\rho}{\sigma}{1} = \tr\rbra*{\frac{\rho\ln\rbra{\rho}+\sigma\ln\rbra{\sigma}}{2} - \frac{\rho+\sigma}{2} \ln\rbra*{\frac{\rho+\sigma}{2}}},
\]
where $\dpower{\rho}{\sigma}{0}$ is the Hellinger distance induced by the log Euclidean mean studied in \cite{BGJ19} and $\dpower{\rho}{\sigma}{1}$ is the Jensen-Shannon divergence \cite{MLP05}. 

Despite a lot of work on estimating quantum entropies and distances in the literature \cite{AISW20,BMW16,GL20,SH21,GHS21,CLW20,WZC+23,WZW23,WGL+24,GP22,WZL24,WZ24b,Wan24,LWWZ25,LW25a,WZ24a,CW25,FW25,LW25b,UNWT25,Wan25}, there are only a few results on estimating quantum divergences including those for the von Neumann relative entropy \cite{Hay25,LF25}, Petz--R\'enyi relative entropy \cite{LF25}, and Tsallis relative entropy \cite{BGW25}. 
In this paper, we consider the estimation of a new family of quantum divergences $\dpower{\rho}{\sigma}{p}$. 
A straightforward approach to estimating $\dpower{\rho}{\sigma}{p}$ is to estimate the density operators of both quantum states by quantum state tomography \cite{HHJ+17,OW16,SSW25,PSW25,PSTW25}, which leads to a sample complexity polynomial in the dimension of the quantum states. 
Regarding this, we aim to find dimension-independent approaches so that one can hope for a large speedup in some practical cases. 

\subsection{Main results}

An important case of interest is the low-rank case (cf.\ \cite{GLF10}), where the quantum states are nearly pure, i.e., the rank of the quantum states is much smaller than the dimension. 
Our main result is an efficient estimator for the quantum divergence $\dpower{\rho}{\sigma}{p}$ in the low-rank case, stated as follows. 

\begin{theorem}[Estimator for the quantum divergence with $p$-power means, informal version of \cref{thm:query_estimate_power_divergence,thm:sample_estimate_power_divergence}]
\label{thm:main}
    For any constant $p \in \rbra{0, 1}$, $\dpower{\rho}{\sigma}{p}$ can be estimated to within additive error $\varepsilon$ using $\poly\rbra{r, 1/\varepsilon}$ queries and samples of $\rho$ and $\sigma$, where $r$ is an upper bound on the ranks of $\rho$ and $\sigma$.
\end{theorem}

Our result is obtained by considering the query complexity of this problem when given purified quantum query access (cf.\ \cite{GL20}) to the quantum states based on quantum singular value transformation \cite{GSLW19} and then applying the quantum sample-to-query lifting \cite{WZ25,WZ24b,TWZ25}. 
As a bonus, further speedups can be achieved if the state-preparation circuits of the quantum states are known, since the first part of our approach considers the estimation problem in the purified quantum query access model (see \cref{thm:query_estimate_power_divergence}). 

We further show that estimating $\dpower{\rho}{\sigma}{p}$ is $\mathsf{BQP}$-complete for the low-rank case, where the $\mathsf{BQP}$-containment is due to \cref{thm:main} and the $\mathsf{BQP}$-hardness holds even if both quantum states are pure (i.e., of rank $1$).

\begin{theorem}[Computational hardness, informal version of \cref{thm:completeness_lowrank_dpower}]\label{thm:bqp-main}
    The decision version of estimating $\dpower{\rho}{\sigma}{p}$ for $n$-qubit quantum states $\rho$ and $\sigma$ of rank $\poly\rbra{n}$ to within additive error $1/\poly\rbra{n}$ is $\mathsf{BQP}$-complete.
\end{theorem}

\cref{thm:bqp-main} is a complement to the literature on the computational complexity of quantum state testing, especially about the $\mathsf{BQP}$-completeness. 
Prior to this work, known $\mathsf{BQP}$-complete problems for quantum state testing include low-rank fidelity estimation \cite{WZC+23,RASW23}, low-rank trace distance estimation \cite{WGL+24,WZ23}, low-rank von Neumann entropy estimation \cite{WGL+24,LW25a}, quantum Tsallis entropy estimation (including purity estimation) \cite{LW25a}, quantum Tsallis relative entropy estimation \cite{BGW25}, low-rank R\'enyi entropy estimation \cite{WGL+24,WZL24,Liu26}, etc.

\subsection{Techniques}

\paragraph{The estimator.}
To estimate $\dpower{\rho}{\sigma}{p}$, the key is to estimate the value of $\tr\rbra*{\rbra*{\frac{\rho^p+\sigma^p}{2}}^{\frac{1}{p}}}$ for $0 < p < 1$.
To achieve this, we utilize the following identity:
\begin{equation*}
    \tr\rbra*{\rbra*{\frac{\rho^p+\sigma^p}{2}}^{\frac{1}{p}}} = \tr\rbra*{\rbra*{\frac{\rho^p+\sigma^p}{2}}^{\frac{1}{2p}} \Pi_{\supp\rbra*{\frac{\rho+\sigma}{2}}} \rbra*{\frac{\rho^p+\sigma^p}{2}}^{\frac{1}{2p}}},
\end{equation*}
where $\Pi_{\supp\rbra{A}}$ denotes the projector onto the support of $A$. 
This idea comes from \cite{WGL+24}, where a similar identity was used in a quantum algorithm for trace distance estimation. 
We informally describe our approach as follows. 
\begin{enumerate}
    \item Approximately prepare a subnormalized density operator $\eta \propto \Pi_{\supp\rbra*{\frac{\rho+\sigma}{2}}}$ (by \cref{lem:eigen_threshold_projector}). 
    \item Implement a block-encoding $U$ of $A \coloneqq q_2\rbra*{\frac{q_1\rbra{\rho}+q_1\rbra{\sigma}}{2}} \approx \rbra*{\frac{\rho^p+\sigma^p}{2}}^{\frac{1}{2p}}$ by quantum singular value transformation \cite{GSLW19}, where $q_1\rbra{x} \approx x^{p}$ and $q_2\rbra{x} \approx x^{\frac{1}{2p}}$ are polynomials (given in \cref{lem:pos_approx}). 
    \item Given $\eta$ and $U$, one can obtain the subnormalized density operator $A \eta A$ by the evolution of subnormalized density operators (\cref{lem:evolution_density_operator}). Note that $A = A^\dag$ is Hermitian. 
    \item Finally, estimate $\tr\rbra{A\eta A}$, which is approximately proportional to $\tr\rbra*{\rbra*{\frac{\rho^p+\sigma^p}{2}}^{\frac{1}{p}}}$, by quantum amplitude estimation \cite{BHMT02}. 
\end{enumerate}
In addition to the standard error analysis, it turns out that we face the following difficulty.
When bounding the bias of the estimator, one has to use $\Abs{q_1\rbra{\rho} - \rho^p}$ and $\Abs{q_1\rbra{\sigma} - \sigma^p}$ (which are easily obtained using the properties of polynomial approximation) to give an upper bound on the following quantity (in \cref{prop:p1_approx})
\begin{equation} \label{eq:diff-sp1}
    \abs*{\tr\rbra*{\rbra*{\frac{q_1\rbra{\rho}+q_1\rbra{\sigma}}{2}}^{\frac{1}{2p}} \eta \rbra*{\frac{q_1\rbra{\rho}+q_1\rbra{\sigma}}{2}}^{\frac{1}{2p}}} - \tr\rbra*{\rbra*{\frac{\rho^p+\sigma^p}{2}}^{\frac{1}{2p}} \eta \rbra*{\frac{\rho^p+\sigma^p}{2}}^{\frac{1}{2p}}}},
\end{equation}
for which the matrix H\"older inequality (\cref{fact:holder}) simply gives 
\begin{equation} \label{eq:diff-sp1-2}
    \eqref{eq:diff-sp1} \leq \Abs*{\rbra*{\frac{q_1\rbra{\rho}+q_1\rbra{\sigma}}{2}}^{\frac{1}{p}} - \rbra*{\frac{\rho^p+\sigma^p}{2}}^{\frac{1}{p}} }.
\end{equation}
To resolve this issue, we provide an inequality for the powers of two positive semi-definite matrices $A$ and $B$ of rank $r$ (in \cref{lem:norm-bound}) that may be of independent interest:
\[
\Abs*{A^{\alpha} - B^{\alpha}} \le 2r \alpha \rbra{\Abs{A}+\Abs{B}}^{\alpha-1} \Abs*{A-B}
\]
for any real $\alpha > 1$, which is based on the Ando--Horn--Johnson inequality \cite{AHJ87} and the Dalecki{\u{\i}}--Kre{\u{\i}}n theorem \cite{DK65}. 
With this matrix inequality, we immediately have 
\begin{equation}
    \eqref{eq:diff-sp1-2} \leq \frac{2^{\frac{1}{p}-1}}{p} r \rbra*{\Abs*{q_1\rbra{\rho}-\rho^{p}} + \Abs*{q_1\rbra{\sigma}-\sigma^{p}}}.
\end{equation}
Here, to ensure that $q_1\rbra{\rho}+q_1\rbra{\sigma}$ is positive semi-definite, we provide an approximation polynomial $q_1\rbra{x}$ for $x^p$ based on the efficiently computable approximation polynomial in \cite{LW25a} (see \cref{lem:modified_pos_approx}) such that $q_1\rbra{x} \geq 0$ for $x \in \sbra{0, 1}$ which ensures $q_1\rbra{\rho} \geq 0$. 

\paragraph{Computational hardness.}
To show the $\mathsf{BQP}$-hardness of estimating $\dpower{\rho}{\sigma}{p}$, we reduce the pure-state trace distance estimation (which is known to be $\mathsf{BQP}$-complete \cite{WZ23}) to it. 
To this end, we need to bound $\dpower{\rho}{\sigma}{p}$ using trace distance from both sides. 
For our purpose, we provide an inequality when $\rho$ and $\sigma$ are pure states (in \cref{lem:dtr_vs_dpower}): 
\begin{equation*}
    \frac{\frac{1}{p}-1}{2^{\frac{1}{p}+1}}\dtr{\rho}{\sigma}{2} \leq \dpower{\rho}{\sigma}{p} \leq \frac{1}{2p}\dtr{\rho}{\sigma}{2}.
\end{equation*}
With this inequality, for pure states $\rho$ and $\sigma$, we immediately have
\begin{equation*}
    \begin{matrix}
        \dtr{\rho}{\sigma}{} \geq a & \Longrightarrow & \dpower{\rho}{\sigma}{p} \geq \frac{\frac{1}{p}-1}{2^{\frac{1}{p}+1}}a^2, \\
        \dtr{\rho}{\sigma}{} \leq b & \Longrightarrow & \dpower{\rho}{\sigma}{p} \leq \frac{1}{2p}b^2,
    \end{matrix}
\end{equation*}
which presents a reduction from pure-state trace distance estimation to pure-state $\dpower{\rho}{\sigma}{p}$ estimation in terms of the parameters $a$ and $b$, thereby showing the $\mathsf{BQP}$-hardness of estimating $\dpower{\rho}{\sigma}{p}$ for certain choices of $a$ and $b$. 

\subsection{Discussion}

In this paper, we consider the computational complexity of estimating the quantum divergence with $p$-power means, $\dpower{\rho}{\sigma}{p}$. 
This type of quantum divergence was recently investigated in \cite{LL21}, which may have potential applications in this line of research. 
We present efficient quantum algorithms for estimating $\dpower{\rho}{\sigma}{p}$ and show that this task is $\mathsf{BQP}$-hard. 
Our results complement the study of quantum divergences with $p$-power means. 
An immediate direction for future research is to consider the estimation of other quantum divergences. 
For reference, the existing work \cite{Hay25,LF25,BGW25} has considered the estimation of von Neumann, Petz--R\'enyi, and Tsallis relative entropies. 
Furthermore, an interesting point is to find practical applications of quantum divergence estimation. 
A possible application is in quantum Boltzmann machine learning \cite{Wil25}, where Tsallis relative entropy was used. 

\section{Preliminaries}

\subsection{Basic notation and facts}
The sets of integers, real numbers, and complex numbers are denoted by
$\Z$, $\R$, and $\C$, respectively. The set of $N$ by $N$ complex matrices is denoted by
$\C^{N \times N}$.
We write $\R\sbra{x}$ for the ring of polynomials in $x$ with real
coefficients.
In this paper, we use operators and matrices interchangeably. 
For a linear operator $A$ on a finite-dimensional Hilbert space,
write $A^\dagger$ for its adjoint and $\rank\rbra{A}$ for its rank, and
define $\abs{A}\coloneqq\sqrt{A^\dagger A}$. For
$s\in\interval[open right]{1}{+\infty}$, define its Schatten $s$-norm by
$\Abs{A}_s\coloneqq
\rbra{\tr\rbra{\abs{A}^s}}^{1/s}$.
Let $\Abs{A}$ denote the operator norm of $A$. 
We write $I$ for the identity operator on the relevant space.
For a nonnegative integer $j$, $I_j$ denotes the identity operator on $j$ qubits and $\ket{0}_j=\ket{0}^{\otimes j}$.
For a linear map $\Phi$ between operator spaces, $\Abs{\Phi}_\diamond$ denotes its diamond norm.
A density operator is a positive semidefinite operator with unit trace.
For an interval $J\subseteq\R$, a function $f\colon J\to\R$, and a
Hermitian operator $A$ whose spectrum is contained in $J$,
choose an orthonormal eigenbasis with
$A=\sum_j\lambda_j\ketbra{\psi_j}{\psi_j}$.
Define $f\rbra{A}=\sum_j f\rbra{\lambda_j}\ketbra{\psi_j}{\psi_j}$.
For $\theta\geq0$ define its $\theta$-support by
$\supp_\theta\rbra{A}
\coloneqq \spanspace\cbra{\ket{\psi_j}\colon\abs{\lambda_j}>\theta}$.
In particular, $\supp\rbra{A}\coloneqq\supp_0\rbra{A}$ is the support of $A$. We write $\Pi_{\supp_\theta\rbra{A}}$ for the orthogonal projector onto $\supp_\theta\rbra{A}$.
We need the following operator--trace norm inequality.
\begin{theorem}[Operator--trace norm H\"older inequality
    {\cite[Theorem 2]{Bau11}}]\label{fact:holder}
    For every positive integer $n$ and all matrices
    $A,B\in\C^{n\times n}$,
    $\abs*{\tr\rbra*{A^\dagger B}} \leq\Abs*{A}\Abs*{B}_1$.
\end{theorem}

The following two facts will be used.

\begin{fact}[Bernoulli inequality]\label{fact:Bernoulli_inequality}
    For a real number $\alpha \geq 1$ and $x \geq -1$,
    $\rbra{1+x}^\alpha \geq 1 + \alpha x$.
\end{fact}

\begin{fact}[Lagrange mean value theorem]\label{fact:mean_value_theorem}
    Let $a,b\in\R$ satisfy $a<b$, and let
    $f\colon\interval{a}{b}\to\R$ be continuous on
    $\interval{a}{b}$ and differentiable on $\interval[open]{a}{b}$.
    Then there exists
    $\xi\in\interval[open]{a}{b}$ such that
    $f^\prime\rbra{\xi}=\frac{f\rbra{b}-f\rbra{a}}{b-a}$.
\end{fact}

\subsection{Matrix analysis}

We first introduce the convexity of trace functions of matrices. 

\begin{theorem}[Convexity of trace functions {\cite[Theorem 2.10]{Car10}}]\label{thm:convexity_operator_function}
    Let $f \colon \R \to \R$ be a continuous function. If $f\rbra{x}$ is (strictly) convex, then function $F\rbra{A} = \tr\rbra{f\rbra{A}}$ is (strictly) convex on Hermitian operators, i.e., $F\rbra{\rbra{1-t}A + tB} \leq \rbra{1-t}F\rbra{A} + tF\rbra{B}$ for any $A, B$ are Hermitian.
\end{theorem}

Let $f \colon \R \to \R$ be a function. 
For Hermitian matrices $A$ and $B$, we use $A \leq B$ to denote that $A$ is less than $B$ in Lowner order, i.e., $B-A$ is positive semi-definite. A function $f$ is said to be operator monotone if for any positive integer $n$, $A \le B$ implies $f\rbra{A} \le f\rbra{B}$, for any $n$ by $n$ Hermitian matrices $A$ and $B$.
A function $f$ is said to be operator convex if for any positive integer $n$, real number $t \in \interval{0}{1}$, there exists
\begin{equation*}
    f\rbra*{\rbra*{1-t} A + t B} \le \rbra*{1-t}f\rbra*{A} + t f\rbra*{B}
\end{equation*}
for any $n$ by $n$ Hermitian matrices $A$ and  $B$. A function $f$ is called operator concave if the function $-f$ is operator convex.
As an example of an operator monotone function, we need the L\"owner--Heinz theorem \cite{Low34,Hei51}.
We use the version in \cite{Bha97}.

\begin{theorem}[L\"owner--Heinz theorem {\cite[Theorem V.1.9]{Bha97}}]\label{thm:x^r_monotone}
    The function $f\rbra{x} = x^\alpha$ is operator monotone on $\interval[open right]{0}{\infty}$ for $0 \leq \alpha \leq 1$.
\end{theorem}

The following theorem reveals the equivalence between operator monotone and operator concave functions.

\begin{theorem}[{\cite[Theorem V.2.5]{Bha97}}]\label{thm:monotone_concavity}
    Let $f\colon \R_+ \to \R_+$ be a continuous function. Then $f$ is operator concave if and only if $f$ is operator monotone.
\end{theorem}

For operator convex functions, we need the following good property \cite{Dav57,Cho74}. We use the version in \cite{Bha97}.

\begin{theorem}[Choi--Davis--Jensen inequality {\cite[Theorem V.2.3]{Bha97}}]\label{thm:choi_davis_jensen}
    Let $I$ be an interval containing $0$ and let $f$ be a real function on $I$. Then the following conditions are equivalent:
    \begin{itemize}
        \item $f\rbra{K_1 A K_1^\dagger + K_2 B K_2^\dagger} \leq K_1 f\rbra{A} K_1^\dagger + K_2 f\rbra{B} K_2^\dagger$ for all operators $K_1$, $K_2$ such that $K_1 K_1^\dagger + K_2 K_2^\dagger \leq I$ and for all Hermitian $A$, $B$ with spectrum in $I$;
        \item $f$ is operator convex on $I$ and $f\rbra{0} \leq 0$.
    \end{itemize}
\end{theorem}

We review the notions and results in matrix differentials (see~\cite[Section V.3]{Bha97}).
Let $f\colon I\to\R$ be continuously differentiable on an interval $I$.
We define its first divided difference $f^{\sbra{1}}$ on $I\times I$ by
\begin{equation*}
    \begin{aligned}
        f^{\sbra{1}}\rbra*{\lambda, \mu} & = \frac{f\rbra*{\lambda}-f\rbra*{\mu}}{\lambda-\mu}, \text{ if } \lambda \ne \mu; \\
        f^{\sbra{1}}\rbra*{\lambda, \lambda} & = f^{\prime}\rbra*{\lambda}.
    \end{aligned}
\end{equation*}
For a Hermitian matrix $A$ with spectrum in $I$, choose an eigenbasis in
which $A=\diag\rbra{\lambda_1,\ldots,\lambda_n}$. In this basis, define
the divided-difference matrix $f^{\sbra{1}}\rbra{A}$ by
$\rbra{f^{\sbra{1}}\rbra{A}}_{ij}
=f^{\sbra{1}}\rbra{\lambda_i,\lambda_j}$.
This matrix notation is always understood in the chosen eigenbasis of $A$.
Let $\calD f\rbra{A}$ denote the Fr\'{e}chet derivative of the matrix
function at $A$. For a Hermitian direction $H$,
\begin{equation*}
    \calD f\rbra*{A}\rbra*{H} = \frac{\mathrm{d}}{\mathrm{d}t}\bigg\vert_{t=0} f\rbra*{A+tH}.
\end{equation*}

We need the following theorem for the derivative of a matrix function, given in
\cite{DK65}. 
Here, we use the version in \cite{Bha97}. 

\begin{theorem}[Dalecki{\u{\i}}--Kre{\u{\i}}n theorem {\cite[Theorem V.3.3]{Bha97}}]
    \label{thm:differential-matrix-function}
    Let $f\colon I\to\R$ be continuously differentiable on an open
    interval $I$. Let $A$ and $H$ be Hermitian matrices of the same size,
    with the spectrum of $A$ contained in $I$. In an eigenbasis of $A$,
    \begin{equation*}
        \calD f\rbra{A}\rbra{H}
        =f^{\sbra{1}}\rbra{A}\circ H,
    \end{equation*}
    where all matrices are represented in this basis and $\circ$ denotes
    the Schur product (entry-wise product).
\end{theorem}

We also need the following basic inequality for the Schur product of Hermitian matrices given in \cite{AHJ87}. 
Here, we use the version from \cite{HJ91}. 

\begin{theorem}[Ando--Horn--Johnson inequality {\cite[Theorem 5.5.1]{HJ91}}]
    \label{prop:multiplicative-norm-bound-schur}
    Let $A$ and $B$ be Hermitian operators on $\calH$. Then
    \begin{equation*}
        \norm{A\circ B} \le \norm{A} \norm{B}, 
    \end{equation*}
    where $\circ$ denotes the Schur product (entry-wise product).
\end{theorem}

\subsection{Model of computation}\label{sec:computational_model}

The standard quantum circuit model is used in this work. 

\textbf{Quantum query complexity.}
Given a set of quantum unitary oracles $\cbra{U_1,U_2,\dots,U_k}$, a quantum query algorithm $\calA^{U_1,\ldots,U_k}$ can be described by a quantum circuit that prepares the state
\begin{equation*}
    A^{\rbra*{T}} U^{\rbra*{T}} \dots A^{\rbra*{t}} U^{\rbra*{t}} \dots A^{\rbra*{1}} U^{\rbra*{1}} A^{\rbra*{0}} \ket{0},
\end{equation*}
where each $U^{\rbra*{t}}$ is a query to (controlled-)$U_j$ or (controlled-)$U_j^\dagger$ for some $1 \leq j \leq k$ and each $A^{\rbra*{t}}$, for $0 \leq t \leq T$, is a unitary operator composed of one- and two-qubit quantum gates, which is independent of $\cbra{U_1,U_2,\ldots,U_k}$. The algorithm then performs a final measurement. The algorithm $\calA^{U_1,\ldots,U_k}$ is said to take $T$ queries to $\cbra{U_1,U_2,\dots,U_k}$. Therefore, the query complexity of $\calA^{U_1,\ldots,U_k}$ is $T$ and its time complexity is the sum of $T$ plus the number of one- and two-qubit gates to implement $A^{(0)},\ldots,A^{(T)}$.
In this work, we consider the following oracle access model, called \textit{purified quantum query access} \cite{GL20}.

\begin{definition}[Purified quantum query access]
\label{def:purified-quantum-query-access}
Let $\rho$ be an unknown $n$-qubit quantum state. A $\rbra{n+a}$-qubit unitary operator $\calO_\rho$ is said to be a purified quantum query access oracle for $\rho$ if
\begin{equation*}
    \ket{\phi} = \calO_\rho \ket{0}_n \ket{0}_a,
\end{equation*}  
where $\ket{\phi}$ is a purification of $\rho$, i.e., $\rho = \tr_a\rbra{\ketbra{\phi}{\phi}}$, and $\tr_a$ denotes the partial trace over the $a$-qubit ancillary register.
\end{definition}

\textbf{Quantum sample complexity.}
For a quantum algorithm $\calA^S$ with samples of quantum states as input, we allow the algorithm in the form $\calE\rbra{\bigotimes_{i}\rho_i^{\otimes k_i}}$, where $k_i$ is the number of independent copies of each $\rho_i$ and $\calE$ is a quantum channel independent of the unknown input states, implemented by a circuit using one- and two-qubit gates, ancillary qubits, and measurements. Therefore, the sample complexity of $\calA^S$ is $Q=\sum_i k_i$ and its time complexity is $Q$ plus the number of one- and two-qubit gates to implement the quantum channel $\calE$.

\subsection{Block-encoding}

A block-encoding represents a scaled approximation of a linear operator in
the top-left block of a possibly larger operator. The encoding operator
need not be unitary. Below we use both unitary operators and density
operators.

\begin{definition}[Block-encoding]\label{def:block-encoding}
    Suppose that $A$ is an $n$-qubit linear operator. For real numbers $\alpha>0$ and $\eps\geq0$ and a nonnegative integer $a$, an $\rbra{n+a}$-qubit operator $B$ is said to be an $\rbra{\alpha,a,\eps}$-block-encoding of $A$ if
    \begin{equation*}
        \Abs*{\alpha\bra{0}_a B\ket{0}_a - A} \leq \eps.
    \end{equation*}
\end{definition}

We define preparation of a subnormalized density operator using
a density-operator block-encoding.

\begin{definition}[Subnormalized density operator]\label{def:subnormalized-density-operator}
    A positive semi-definite operator $A$ is called a subnormalized density operator if $0 \leq \tr\rbra{A} \leq 1$. Moreover, the density operator is a subnormalized density operator with unit trace. An $\rbra{n+a+b}$-qubit unitary operator $U$ is said to prepare the $n$-qubit subnormalized density operator $A$ if $U$ can prepare the purification $\ket{\sigma}=U\ket{0}_{n+a+b}$ of $\rbra{n+a}$-qubit density operator i.e. $\tr_b\rbra{\ketbra{\sigma}{\sigma}}$, which is a $\rbra{1,a,0}$-block-encoding of $A$. Moreover, $a=0$ when $A$ is a density operator.
\end{definition}

Given a subnormalized density operator $A$ prepared by a unitary $U$, we use
the following construction to obtain a block-encoding
$\widetilde{U}$ of $A$. This technique was introduced in \cite{LC19} and
later generalized to subnormalized density operators in
\cite{GSLW19,vAG19}.

\begin{lemma}[Block-encoding of a subnormalized density operator \cite{LC19,GSLW19,vAG19}]\label{lem:block-encode-subnormalized-density-operator}
    Suppose $U$ is an $\rbra{n+a}$-qubit unitary operator that prepares $n$-qubit subnormalized density operator $A$. Then there is a $\rbra{2n+a}$-qubit unitary operator $\widetilde{U}$ which is an $\rbra{1,n+a,0}$-block-encoding of $A$, using $O\rbra{1}$ query to $U$ and $U^\dagger$ and $O\rbra{a}$ elementary quantum gates.
\end{lemma}

We next define state-preparation pairs for use in the following linear-combination-of-unitaries (LCU) lemma.

\begin{definition}[State-preparation pair {\cite[Definition 51 in the full version]{GSLW19}}]\label{def:state-preparation-pair}
    Let $y=(y_0,\ldots,y_{m-1})\in\C^m$ with $\abs{y}_1\leq\beta$ and $m\leq2^b$. The pair of unitaries $\rbra{P_L,P_R}$ is called a $\rbra{\beta,b,\eps}$-state-preparation-pair for $y$ if $P_L\ket{0}^{\otimes b} = \sum_{j=0}^{2^b-1}c_j\ket{j}$ and $P_R\ket{0}^{\otimes b} = \sum_{j=0}^{2^b-1}d_j\ket{j}$ such that $\sum_{j=0}^{m-1}\abs{\beta\rbra{c_j^\ast d_j}-y_j} \leq \eps$ and $c_j^\ast d_j=0$ for all $j \in \cbra{m,\ldots,2^b-1}$.
\end{definition}

\begin{lemma}[Linear combination of block-encoded matrices {\cite[Lemma 52 in the full version]{GSLW19}}]\label{lem:lcu}
    Let $A = \sum_{j=0}^{m-1} y_j A_j$ is an $s$-qubit operator and $\eps >0$. Suppose that $\rbra{P_L,P_R}$ is a $\rbra{\beta,b,\eps_1}$-state-preparation-pair for $y \in \C^m$, $W=\sum_{j=0}^{m-1}\ketbra{j}{j} \otimes U_j + \sum_{j=m}^{2^b-1}\ketbra{j}{j} \otimes I_a \otimes I_s$ is an $\rbra{s+a+b}$-qubit unitary such that for all $j \in \cbra{0,\ldots,m-1}$, $U_j$ is an $\rbra{\alpha,a,\eps_2}$-block-encoding of $A_j$. Then we can implement a $\widetilde{W}=\mathsf{LCU}\rbra{\cbra{P_L,P_R}, \cbra{U_j}_j}=\rbra{P_L^\dagger \otimes I_a \otimes I_s}W\rbra{P_R \otimes I_a \otimes I_s}$ is a $\rbra{\alpha\beta,a+b,\alpha\eps_1+\alpha\beta\eps_2}$-block-encoding of $A$. Moreover, the unitary $\widetilde{W}$ uses a single use of $W$, $P_L^\dagger$ and $P_R$.
\end{lemma}

Ordinary unitary evolution maps a density operator $A$ to $UAU^\dagger$.
The following lemma shows how to generalize this construction to
conjugation by a possibly nonunitary operator $B$ supplied through a
block-encoding, while allowing $A$ to be subnormalized.

\begin{lemma}[Evolution of a subnormalized density operator {\cite[Lemma II.2]{WGL+24}}]\label{lem:evolution_density_operator}
    Suppose $U$ is an $\rbra{n+a}$-qubit unitary operator that prepares an
    $n$-qubit subnormalized density operator $A$ and $V$ is an
    $\rbra{n+b}$-qubit unitary operator which is a
    $\rbra{1,b,0}$-block-encoding of $B$. Then
    $\widetilde{U}=\mathsf{SubDensityOpEvolution}\rbra{U,V}=\rbra{V \otimes I_a}\rbra{U \otimes I_b}$
    is an $\rbra{n+a+b}$-qubit unitary operator that prepares an
    $n$-qubit subnormalized density operator $B A B^\dagger$.
\end{lemma}

The following lemma uses quantum amplitude estimation \cite{BHMT02} to
estimate the trace of a prepared subnormalized density operator.

\begin{lemma}[Trace estimation of subnormalized density operators {\cite[Lemma II.15]{WGL+24}}]\label{lem:trace_est_subdensity-operator}
    Suppose $U$ is a $\rbra{n+a}$-qubit unitary operator that prepares an $n$-qubit subnormalized density operator $A$, and $C \geq 0$ is a known constant such that $\tr\rbra{A} \leq C$. For every $\eps >0$, there is a quantum algorithm $\mathsf{EstTrSubDensityOp}\rbra{U,C,\eps}$ that estimate $\tr\rbra{A}$ within additive error $\eps$ with $O\rbra{\frac{\sqrt{C}}{\eps}+\frac{1}{\sqrt{\eps}}}$ queries to $U$ and $U^\dagger$.
\end{lemma}

\subsection{Quantum singular value transformation}

Our algorithm uses quantum singular value transformation
(QSVT)~\cite{GSLW19} to implement polynomial eigenvalue transformations
of Hermitian operators.

\begin{theorem}[Polynomial eigenvalue transformation via QSVT {\cite[Theorem 56 in the full version]{GSLW19}}]\label{thm:qsvt}
    Suppose that $U$ is an $\rbra{\alpha,a,\eps}$-block-encoding of Hermitian matrix $A$. Let
    $q\in\R\sbra{x}$ be a $d$-degree polynomial such that $\abs{q\rbra{x}}\leq1/2$ on $\interval{-1}{1}$. For $\delta \geq 0$, there is a quantum circuit $\widetilde{U}=\mathsf{EigenTrans}\rbra{U,q,\delta}$ that is a $\rbra{1,a+2,4d\sqrt{\eps/\alpha}+\delta}$-block-encoding of
    $q\rbra{A/\alpha}$.
    The circuit $\widetilde{U}$ uses $O\rbra{d}$ queries in total to $U$
    and $U^\dagger$, one query to controlled-$U$, and
    $O\rbra{\rbra{a+1}d}$ one- and two-qubit gates. And the description of the circuit $\widetilde{U}$ can be computed classically in time
    $\poly\rbra{d,\log\rbra{1/\delta}}$.
\end{theorem}

As an application of \cref{thm:qsvt}, we need the following result on how to block-encode the eigenvalue threshold projector, which is defined as follows.

\begin{definition}[Eigenvalue threshold projector]
    Given an matrix $A$, $\Pi_{\supp\rbra{A}}$ is said to be the eigenvalue threshold projector of $A$, where $\supp{A}$ is the support of $A$ and $\Pi_{\supp\rbra{A}}$ is the projector into the subspace of $\supp\rbra{A}$.
\end{definition}

\begin{lemma}[Block-encoding of eigenvalue threshold projector {\cite[Lemma II.19]{WGL+24}}]\label{lem:eigen_threshold_projector}
    Let $\delta,\eps \in \interval[open left]{0}{\frac{1}{10}}$ and $32\eps^2 \leq \delta$. Suppose $U$ is an $\rbra{n+a}$-qubit unitary operator that prepares an $n$-qubit subnormalized density operator $A$. For every $\delta^\prime > 0$, there is an unitary operator $\widetilde{U}=\mathsf{EigenThreProj}\rbra{U,\eps,\delta,\delta^\prime}$ such that $\rbra{n+2a+2}$-qubit unitary operator $\widetilde{U}$ prepares an $n$-qubit subnormalized density operator $\widetilde{B}$, which is a $\rbra{1,0,\delta^\prime}$-block-encoding of a positive semidefinite operator $B$ such that
    \begin{equation*}
        \rbra*{\frac{\delta}{4}\rbra*{1-2\eps}-\delta^{1/2}\eps} \Pi_{\supp_{2\delta}\rbra{A}} \leq B \leq \rbra*{\frac{\delta}{4}+\eps^2+\delta^{1/2}\eps}\Pi_{\supp\rbra{A}}.
    \end{equation*}
    Both $\widetilde{B}$ and $B$ have rank at most $\rank\rbra{A}$.
    Moreover, $\widetilde{U}$ uses $O\rbra{d}$ queries to $U$ and $U^\dagger$, one query to controlled-$U$ and $O\rbra{\rbra{n+a}d}$ elementary quantum gates, where $d=O\rbra{\frac{1}{\delta}\log\rbra{\frac{1}{\eps}}}$. The description of $\widetilde{U}$ can be computed in $\poly\rbra{d,\log\rbra{\frac{1}{\delta^\prime}}}$ time classically.
\end{lemma}

\subsection{Quantum samplizer}\label{sec:quantum_samplizer}

The quantum samplizer \cite{WZ25,WZ24b} converts a quantum query algorithm into one
that uses only independent copies of the input state.
Here, we use the version of multi-samplizer extended in \cite{BGW25}.

\begin{definition}[Quantum multi-samplizer {\cite[Definition~2.22]{BGW25}}]
\label{def:multi-samplizer}
    Fix positive integers $k$ and $m$. A $k$-samplizer, denoted by
    $\mathsf{Samplize}_*\langle*\rangle\sbra{*}$, converts a quantum query
    algorithm into a quantum algorithm with sample access as follows.
    For every positive integer $n$, every precision $\zeta>0$, every
    quantum query algorithm
    $\calA^{U_1,\ldots,U_k}$ with query access to unitary oracles
    $U_1,\ldots,U_k$ acting on $n+m$ qubits each, and all
    $n$-qubit density operators
    $\rho_1,\ldots,\rho_k$, the quantum channel
    $\mathsf{Samplize}_\zeta\langle\calA^{U_1,\ldots,U_k}\rangle
    \sbra{\rho_1,\ldots,\rho_k}$ uses only independent copies of
    $\rho_1,\ldots,\rho_k$, and there exist
    unitary $\rbra{1,m,0}$-block-encodings $U_{\rho_1},\ldots,U_{\rho_k}$
    of $\rho_1/2,\ldots,\rho_k/2$, respectively, such that
    \begin{equation*}
        \Abs*{\mathsf{Samplize}_{\zeta}
        \langle\calA^{U_1,\ldots,U_k}\rangle
        \sbra{\rho_1,\ldots,\rho_k}
        -\calA^{U_{\rho_1},\ldots,U_{\rho_k}}}_\diamond
        \leq\zeta,
    \end{equation*}
    where the circuit $\calA^{U_{\rho_1},\ldots,U_{\rho_k}}$ is identified
    with the quantum channel it induces.
\end{definition}

\begin{lemma}[Quantum multi-samplizer
{\cite[Theorem~2.23]{BGW25}}]
\label{lem:multi-samplizer}
    Fix positive integers $n$ and $k$. There is a $k$-samplizer with
    $m=4$ such that the following holds.
    Let $\rho_1,\ldots,\rho_k$ be
    $n$-qubit density operators, and let
    $\zeta\in\interval[open]{0}{1}$. Suppose a quantum query algorithm
    $\calA^{U_1,\ldots,U_k}$ uses $Q_j$ queries to $U_j$ for every $1 \leq j \leq k$. The $\mathsf{Samplize}_\zeta\langle\calA^{U_1,\ldots,U_k}\rangle
    \sbra{\rho_1,\ldots,\rho_k}$ uses
    \begin{equation*}
        O\rbra*{\frac{Q_jQ}{\zeta}
        \log^2\rbra*{\frac{Q}{\zeta}}}
    \end{equation*}
    copies of $\rho_j$ for each $j$, where $Q = \sum_{j=1}^k Q_j$.
    
    Moreover, if $\calA^{U_1,\ldots,U_k}$ uses $T$ gates, then $\mathsf{Samplize}_\zeta\langle\calA^{U_1,\ldots,U_k}\rangle
    \sbra{\rho_1,\ldots,\rho_k}$ use 
    \begin{equation*}
        T+O\rbra*{\frac{Q^2n}{\zeta}
        \log^2\rbra*{\frac{Q}{\zeta}}}
    \end{equation*}
    gates.
\end{lemma}

\subsection{Polynomial approximation}

For QSVT, we need a polynomial that approximates a real-valued extension
of a positive power function while remaining uniformly bounded on
$\interval{-1}{1}$. The following lemma provides such a polynomial.

\begin{lemma}[Polynomial approximation of positive power functions {\cite[Lemma 3.1]{LW25a}}]\label{lem:pos_approx}
    Let $\eps \in \interval[open]{0}{1/2}$. And let $k$ be a positive constant integer and $\gamma$ be a constant real number where $\gamma \in \interval[open]{-1}{1}$. For any function $f\rbra{x} \coloneqq \frac{1}{2}x^{k-1}\abs{x}^{1+\gamma}$, there exists a polynomial $q \in \R\sbra{x}$ such that
    \begin{itemize}
        \item $\abs{q\rbra{x}-f\rbra{x}} \leq \eps$ for $x \in \interval{-1}{1}$, and
        \item $\abs{q\rbra{x}} \leq 1$ for $x \in \interval{-1}{1}$.
    \end{itemize}
    Moreover, the degree of the polynomial $q\rbra{x}$ is $O\rbra{\rbra{\frac{1}{\eps}}^{\frac{1}{k+\gamma}}}$, and the coefficients of the polynomial $q\rbra{x}$ can be computed in classical polynomial time.
\end{lemma}

\subsection{Closeness testing of quantum states}

We first recall the definition of trace distance.

\begin{definition}[Trace distance]
    For density operators $\rho,\sigma$ on the same finite-dimensional
    Hilbert space, their trace distance is
    \begin{align*}
        \dtr{\rho}{\sigma}{}\coloneqq\frac{1}{2}\Abs{\rho-\sigma}_1.
    \end{align*}
\end{definition}

We define the following family $\PureQSD\sbra{a,b}$ of promise problems,
parameterized by threshold functions $a$ and $b$, for testing the trace
distance between two pure states.

\begin{definition}[$\PureQSD$]
    Let $Q_\phi$ and $Q_\psi$ be two quantum circuits with $m\rbra{n}$-qubit input and $n$-qubit output, where $m\rbra{n}$ is a polynomial in $n$. 
    Let $\ket{\phi}$ and $\ket{\psi}$ be $n$-qubit pure quantum states obtained by performing $Q_\phi$ and $Q_\psi$ on input state $\ket{0}^{\otimes m\rbra{n}}$. 
    Let $a\rbra{n}$ and $b\rbra{n}$ be efficiently computable functions such that $0 \leq b\rbra{n} < a\rbra{n} \leq 1$. The problem $\PureQSD\sbra{a,b}$ is to decide whether:
    \begin{itemize}
        \item \textup{(Yes)} $\dtr{\ketbra{\phi}{\phi}}{\ketbra{\psi}{\psi}}{} \geq a\rbra{n}$;
        \item \textup{(No)} $\dtr{\ketbra{\phi}{\phi}}{\ketbra{\psi}{\psi}}{} \leq b\rbra{n}$.
    \end{itemize}    
\end{definition}

We will use the following explicit hardness consequence of
\cite[Lemma 2.17]{LW25a}. See also \cite{RASW23,WZ23} for related hardness
results on pure-state closeness testing.

\begin{lemma}[$\BQP$-hardness of $\PureQSD$ {\cite[Lemma 2.17]{LW25a}}]\label{lem:BQP-hard}
    Let $a\rbra{n}$ and $b\rbra{n}$ be efficiently computable functions such that $0 \leq b\rbra{n} < a\rbra{n} \leq 1$ and $a\rbra{n}-b\rbra{n} \geq 1/\poly\rbra{n}$.
    Then, $\PureQSD\sbra*{a, b}$ is $\BQP$-hard when $a\rbra{n} \leq 1-2^{-n-1}$ and $b\rbra{n} \geq 2^{-n-1}$.
\end{lemma}

\section{Algorithms}

This section proves
\cref{thm:query_estimate_power_divergence,thm:sample_estimate_power_divergence}.
It suffices to estimate $\power{\rho}{\sigma}{p}$ to additive error
$\eps$, since subtracting this estimate from one gives an estimate
of $\dpower{\rho}{\sigma}{p}$ with the same accuracy.

We first establish the polynomial approximation and matrix-power
bounds in
\cref{sec:modified_pos_approx,sec:bounds_on_matrix_power_functions}.
\Cref{sec:construct_unitary} gives the shared construction and
specifies the requirements on its eigenvalue-threshold operator;
\cref{sec:error_analysis} bounds the resulting approximation and
implementation errors.
\Cref{sec:query_algorithm,sec:sample_algorithm} realize this
construction in the purified query and sample access models,
respectively, verify these requirements, and account for the
additional estimation and simulation errors to obtain the
resource bounds.

\subsection{Modified approximating polynomial}\label{sec:modified_pos_approx}
We modify the polynomial in \cref{lem:pos_approx} to make it nonnegative on $\interval{0}{1}$.
The reason comes from the fact that $A^{\alpha}$ might not be well-defined when $\alpha>0$ is not an integer, if $A$ is just Hermitian but not positive semi-definite (see \cref{lem:norm-bound}).

\begin{lemma}\label{lem:modified_pos_approx}
    Let $\eps \in \interval[open]{0}{\frac{1}{2}}$. And let $k$ be a fixed positive constant integer and $\gamma$ be a constant real number where $\gamma \in \interval[open]{-1}{1}$. For any function $f\rbra{x} \coloneqq \frac{1}{4}x^{k-1}\abs{x}^{1+\gamma}$, there exists a polynomial $q \in \R\sbra{x}$ such that
    \begin{itemize}
        \item $\abs*{q\rbra{x}-f\rbra{x}} \leq \eps$ for $x \in \interval{-1}{1}$, 
        \item $\abs*{q\rbra{x}} \leq 1$ for $x \in \interval{-1}{1}$, and
        \item $0 \leq q\rbra{x} \leq 1$ for $x \in \interval{0}{1}$.
    \end{itemize}
    Moreover, the degree of the polynomial $q\rbra{x}$ is $O\rbra{\rbra{\frac{1}{\eps}}^{\frac{1}{k+\gamma}}}$, and the coefficients of the polynomial $q\rbra{x}$ can be computed in classical polynomial time.
\end{lemma}

\begin{proof}
    Suppose $q^\prime$ is the polynomial in \cref{lem:pos_approx} with $\abs{q^\prime\rbra{x}-2f\rbra{x}} \leq \eps^\prime$ and $\abs{q^\prime\rbra{x}} \leq 1$ for $x \in \interval{-1}{1}$, where $\eps^\prime \in \interval[open]{0}{\frac{1}{2}}$.
    Let $q\rbra{x} \coloneqq \rbra{q^\prime\rbra{x}+\eps^\prime}/2$ and then
    \begin{equation*}
        q\rbra*{x} \leq \frac{\eps^\prime+1}{2} \leq \frac{3}{4} \leq 1 \quad \text{and} \quad q\rbra*{x} \geq \frac{\eps^\prime-1}{2} \geq - \frac{1}{2} \geq -1
    \end{equation*}
    for $x \in \interval{-1}{1}$. When $x \in \interval{0}{1}$,
    \begin{equation*}
        q^\prime\rbra*{x} \geq 2f\rbra*{x}-\eps^\prime \geq - \eps^\prime
    \end{equation*}
    where $f\rbra{x} \geq 0$ for $x \in \interval{0}{1}$ by the definition.
    Therefore, for $x \in \interval{0}{1}$,
    \begin{equation*}
        q\rbra{x} \geq \frac{-\eps^\prime+\eps^\prime}{2} \geq 0. 
    \end{equation*}

    Also 
    \begin{equation*}
        \abs*{q\rbra*{x}-f\rbra*{x}} = \abs*{\frac{q^\prime\rbra*{x} - 2f\rbra*{x} + \eps^\prime}{2}} \leq \eps^\prime \eqcolon \eps        
    \end{equation*}
    where $\eps \in \interval[open]{0}{\frac{1}{2}}$. And the degree of the polynomial $q\rbra{x}$ is $O\rbra{\rbra{\frac{1}{\eps^\prime}}^{\frac{1}{k+\gamma}}} = O\rbra{\rbra{\frac{1}{\eps}}^{\frac{1}{k+\gamma}}}$
\end{proof}

\subsection{Bounds on matrix power functions}\label{sec:bounds_on_matrix_power_functions}

We first prove the following lemma.

\begin{lemma}\label{prop:norm-bound-differential}
    Let $\alpha>1$ and $f\rbra{x}=x^\alpha$ for $x\geq0$.
    Let $A$ be a positive semidefinite operator and $H$ a Hermitian
    operator on the same finite-dimensional Hilbert space. Suppose
    $H=PHP$, where $P=\Pi_{\supp\rbra{A}}$. Then
    \begin{equation*}
        \Abs*{f^{\sbra{1}}\rbra{A}\circ H}
        \leq r\alpha M^{\alpha-1}\Abs{H},
    \end{equation*}
    where $r=\rank\rbra{A}$, $M=\Abs{A}$, and $\circ$ denotes the
    Schur product in a basis where $A$ is diagonal.
\end{lemma}

\begin{proof}
    If $A=0$, then $H=0$ and the claim is immediate. Otherwise,
    $H=PHP$ allows us to restrict the Schur product to $\supp\rbra{A}$
    without changing its norm or $\Abs{H}$. In the rest of the proof,
    $A$ and $H$ denote their restrictions to this $r$-dimensional space.
    In an eigenbasis of $A$, write
    $L_{ij}=f^{\sbra{1}}\rbra{\lambda_i,\lambda_j}$, where
    $0<\lambda_i\leq M$. The mean value theorem
    (\cref{fact:mean_value_theorem}) gives
    $\abs{L_{ij}}\leq\alpha M^{\alpha-1}$ for distinct eigenvalues;
    the same bound follows from $f^\prime\rbra{\lambda_i}$ when they
    coincide. Hence $\Abs{L}\leq r\max_{i,j}\abs{L_{ij}}
    \leq r\alpha M^{\alpha-1}$. The Schur-product inequality in
    \cref{prop:multiplicative-norm-bound-schur} now gives
    \begin{equation*}
        \Abs{L\circ H}\leq\Abs{L}\Abs{H}
        \leq r\alpha M^{\alpha-1}\Abs{H}.
    \end{equation*}
\end{proof}

The following lemma can bound the operator norm of fractional matrix power differences.

\begin{lemma}\label{lem:norm-bound}
    For any $\alpha > 1$, and positive semi-definite operators $A$ and $B$, we have
    \begin{equation*}
        \Abs*{A^{\alpha} - B^{\alpha}} \le 2r \alpha M^{\alpha-1} \Abs*{A-B},
    \end{equation*}
    where $M = \Abs{A} + \Abs{B}$, $r = \max\cbra{\rank\rbra{A}, \rank\rbra{B}}$.
\end{lemma}

\begin{proof}
    Set $H=A-B$ and $C_t=\rbra{1-t}B+tA$. For $0<t<1$,
    positivity gives $\ker\rbra{C_t}=\ker\rbra{A}\cap\ker\rbra{B}$,
    and hence $\supp\rbra{C_t}=\supp\rbra{A}+\supp\rbra{B}$.
    Thus $H=P_tHP_t$, where $P_t=\Pi_{\supp\rbra{C_t}}$, and
    \begin{equation*}
        \rank\rbra{C_t}\leq\rank\rbra{A}+\rank\rbra{B}\leq2r,
        \qquad \Abs{C_t}\leq M.
    \end{equation*}

    The function $g\rbra{x}=\abs{x}^\alpha$ is continuously
    differentiable on $\R$ and agrees with $x^\alpha$ for $x\geq0$.
    Define $F\rbra{t}=g\rbra{C_t}=C_t^\alpha$ for
    $t\in\interval{0}{1}$. By the Dalecki{\u{\i}}--Kre{\u{\i}}n theorem
    (\cref{thm:differential-matrix-function}), $F^\prime\rbra{t}
    =\calD g\rbra{C_t}\rbra{H}$. Computing this derivative in an
    eigenbasis of $C_t$ and applying \cref{prop:norm-bound-differential}
    gives, for $0<t<1$,
    \begin{equation*}
        \Abs*{F^\prime\rbra{t}}
        \leq\rank\rbra{C_t}\alpha\Abs{C_t}^{\alpha-1}\Abs{H}
        \leq2r\alpha M^{\alpha-1}\Abs{A-B}.
    \end{equation*}
    Integrating on compact subintervals of $\interval[open]{0}{1}$
    and using continuity at the endpoints yields
    \begin{equation*}
        \Abs*{A^{\alpha} - B^{\alpha}}
        \leq\int_0^1\Abs*{F^\prime\rbra{t}}\,\mathrm{d}t
        \leq2r\alpha M^{\alpha-1}\Abs{A-B}.
    \end{equation*}
\end{proof}

\begin{lemma}\label{lem:density_operator_cdj}
    For any $0 \leq \alpha \leq 1$, and positive semidefinite operators $A$ and $B$,
    \begin{equation}
        \frac{A^\alpha+B^\alpha}{2} \leq \rbra*{\frac{A+B}{2}}^\alpha.
    \end{equation}
\end{lemma}

\begin{proof}
    The cases $\alpha=0$ and $\alpha=1$ are immediate. For $0<\alpha<1$,
    \cref{thm:x^r_monotone,thm:monotone_concavity} imply that
    $f\rbra{x}=x^\alpha$ is operator concave on
    $\interval[open]{0}{\infty}$, and continuity extends the inequality
    to positive semidefinite operators. Applying operator concavity at
    $t=1/2$ gives
    \begin{equation*}
        \frac{A^\alpha+B^\alpha}{2}
        \leq\rbra*{\frac{A+B}{2}}^\alpha.
    \end{equation*}
\end{proof}

\subsection{Shared construction}\label{sec:construct_unitary}

Fix a polynomial-time computable constant $p\in\interval[open]{0}{1}$, a
scale $c\in\cbra{1,2}$, and positive integers $n$ and $m$. Let
$\rho$ and $\sigma$ be $n$-qubit density operators, and let $r$ be a known
positive integer with
$r\geq\max\cbra{\rank\rbra{\rho},\rank\rbra{\sigma}}$. Set
\begin{equation*}
    \widehat\rho\coloneqq\frac{\rho}{c},\qquad
    \widehat\sigma\coloneqq\frac{\sigma}{c},\qquad
    \mu\coloneqq\frac{\rho+\sigma}{2},\qquad
    \widehat\mu\coloneqq\frac{\widehat\rho+\widehat\sigma}{2}=\frac{\mu}{c}.
\end{equation*}
Let $U_{\widehat\rho}$ and $U_{\widehat\sigma}$ be exact unitary
$\rbra{1,m,0}$-block-encodings of $\widehat\rho$ and $\widehat\sigma$.
Fix accuracy parameters $\eps_{\mathrm{in}},\eps_{\mathrm{out}}\in\interval[open]{0}{1/2}$,
$\delta_{\mathrm{in}},\delta_{\mathrm{out}},\delta_\eta\in\interval[open]{0}{1}$, and
$\theta,\eps_{\mathrm{th}}\in\interval[open left]{0}{1/10}$ with
$32\eps_{\mathrm{th}}^2\leq\theta$ and $\delta_\eta\leq\theta$. The parameter
$\eps_{\mathrm{out}}$ controls the outer polynomial approximation when $1/p\notin2\Z$.
The query algorithm takes $c=1$
(\cref{sec:query_algorithm}), and the sample algorithm takes $c=2$ and
$m=4$ (\cref{sec:sample_algorithm}).

\textbf{Step 1. Construct block-encodings of scaled polynomial
approximations to $\widehat\rho^p$ and $\widehat\sigma^p$.}
Apply \cref{lem:modified_pos_approx} with integer parameter $1$, real
parameter $p-1\in\interval[open]{-1}{0}$, and accuracy $\eps_{\mathrm{in}}$. Its
real-valued target on $\interval{-1}{1}$ is $\abs{x}^p/4$, which agrees
with $x^p/4$ on $\interval{0}{1}$. Thus there exists a polynomial
$q_{\mathrm{in}}\in\R\sbra{x}$ of degree $d_{\mathrm{in}}=O\rbra{\eps_{\mathrm{in}}^{-1/p}}$ such that
\begin{align}
    \abs*{q_{\mathrm{in}}\rbra*{x}-\frac{1}{4}x^p}
    &\leq\eps_{\mathrm{in}}, \quad \text{for all }x\in\interval{0}{1},
    \label{eq:p1_qsvt_approx}\\
    \abs*{q_{\mathrm{in}}\rbra*{x}}
    &\leq1, \quad \text{for all }x\in\interval{-1}{1},
    \notag\\
    0\leq q_{\mathrm{in}}\rbra*{x}
    &\leq1, \quad \text{for all }x\in\interval{0}{1}.
    \label{eq:p1_qsvt_nonnegative}
\end{align}
Since $\widehat\rho$ and $\widehat\sigma$ are positive semidefinite with
trace at most one, their spectra lie in $\interval{0}{1}$ and
$\max\cbra{\Abs{\widehat\rho},\Abs{\widehat\sigma}}\leq1$.
Apply \cref{thm:qsvt} to $U_{\widehat\rho}$ and $U_{\widehat\sigma}$ with
polynomial $q_{\mathrm{in}}/2$ and accuracy $\delta_{\mathrm{in}}$. This gives unitaries $V_\rho$ and
$V_\sigma$ that are $\rbra{1,m+2,\delta_{\mathrm{in}}}$-block-encodings of
$\frac{1}{2}q_{\mathrm{in}}\rbra{\widehat\rho}$ and
$\frac{1}{2}q_{\mathrm{in}}\rbra{\widehat\sigma}$, respectively. By
\cref{eq:p1_qsvt_approx} and functional calculus, these ideal target
matrices approximate $\widehat\rho^p/8$ and $\widehat\sigma^p/8$,
respectively, within operator-norm error $\eps_{\mathrm{in}}/2$.

\textbf{Step 2. Construct a normalization-$2$ block-encoding of $2\nu$.}
Define
\begin{equation*}
    \nu\coloneqq\frac{q_{\mathrm{in}}\rbra{\widehat\rho}+q_{\mathrm{in}}\rbra{\widehat\sigma}}{4},
    \qquad
    \tau\coloneqq\frac{\widehat\rho^p+\widehat\sigma^p}{16}.
\end{equation*}
Let $H$ denote the Hadamard gate.
Since $\rbra{H,H}$ is a $\rbra{2,1,0}$-state-preparation-pair for
the coefficient vector $\rbra{1,1}$, \cref{lem:lcu} constructs $U_\nu$, a
$\rbra{2,m+3,2\delta_{\mathrm{in}}}$-block-encoding of
$2\nu=\rbra{q_{\mathrm{in}}\rbra{\widehat\rho}+q_{\mathrm{in}}\rbra{\widehat\sigma}}/2$.
By \cref{eq:p1_qsvt_nonnegative}, $0\leq2\nu\leq I_n$, and hence
$\Abs{2\nu}\leq1\leq2$, which verifies the norm hypothesis of
\cref{thm:qsvt} for Step~3.
Since $U_\nu$ has normalization $2$, the normalized matrix argument in the
subsequent QSVT applied to $U_\nu$ is exactly $\nu$, which, by
\cref{eq:p1_qsvt_approx}, satisfies $\Abs{\nu-\tau}\leq\eps_{\mathrm{in}}/2$.

\textbf{Step 3. Construct a block-encoding of a scaled approximation to
$\tau^{1/(2p)}$.}
We consider two cases, according to whether $\frac{1}{p}\in2\Z$. We call
Case~3.1 the exact branch and Case~3.2 the approximate branch. In both
cases, write $q_{\mathrm{out}}$ for the outer polynomial and
$d_{\mathrm{out}}=\deg\rbra{q_{\mathrm{out}}}$ for its degree.

\textbf{Case 3.1.}
When $\frac{1}{p}\in2\Z$, define the polynomial
$q_{\mathrm{out}}\rbra{x}\coloneqq x^{1/(2p)}/4$ of degree
$d_{\mathrm{out}}=1/(2p)$. It satisfies
$\abs{q_{\mathrm{out}}\rbra{x}}\leq1$ on $\interval{-1}{1}$ and
$q_{\mathrm{out}}\rbra{x}\geq0$ on $\interval{0}{1}$.

\textbf{Case 3.2.}
When $\frac{1}{p} \notin 2\Z$, we have
$\frac{1}{2p}\notin\Z$. Apply \cref{lem:modified_pos_approx} with integer
parameter $\lceil 1/(2p)\rceil$, real parameter
$1/(2p)-\lceil 1/(2p)\rceil\in\interval[open]{-1}{0}$, and accuracy
$\eps_{\mathrm{out}}$. The lemma's real-valued target on $\interval{-1}{1}$ agrees with
$x^{1/(2p)}/4$ on $\interval{0}{1}$. Thus there exists a polynomial
$q_{\mathrm{out}}\in\R\sbra{x}$ of degree $d_{\mathrm{out}}=O\rbra{\eps_{\mathrm{out}}^{-2p}}$ such that
\begin{align}
    \abs*{q_{\mathrm{out}}\rbra{x}-\frac{1}{4}x^{1/(2p)}}
    &\leq\eps_{\mathrm{out}}, \quad \text{for all }x\in\interval{0}{1},
    \label{eq:p2_qsvt_approx}\\
    \abs*{q_{\mathrm{out}}\rbra{x}}
    &\leq1, \quad \text{for all }x\in\interval{-1}{1},
    \notag
\end{align}

\textbf{Common implementation.}
In either branch, apply \cref{thm:qsvt} to $U_\nu$ with polynomial
$q_{\mathrm{out}}/2$ and accuracy $\delta_{\mathrm{out}}$. This gives a unitary $U_D$
that is a $\rbra{1,m+5,4d_{\mathrm{out}}\sqrt{\delta_{\mathrm{in}}}+\delta_{\mathrm{out}}}$-block-encoding of
$D\coloneqq\frac12 q_{\mathrm{out}}\rbra{\nu}$.
In the exact branch, $D=\nu^{1/(2p)}/8$; in the approximate branch,
\cref{eq:p2_qsvt_approx} gives
$\Abs{D-\nu^{1/(2p)}/8}\leq\eps_{\mathrm{out}}/2$.
Let $\widetilde D$ be the actual top-left block of $U_D$.
The spectrum of $\nu$ lies in $\interval{0}{1/2}$, and
$q_{\mathrm{out}}$ is bounded in absolute value by one on
$\interval{0}{1}$. Thus both branches satisfy
\begin{equation}\label{eq:outer_block_error}
    \Abs*{\widetilde D-D}\leq \delta_D\coloneqq4d_{\mathrm{out}}\sqrt{\delta_{\mathrm{in}}}
    +\delta_{\mathrm{out}},\qquad
    \Abs*{D}\leq\frac{1}{2},\qquad
    \Abs*{\widetilde D}\leq1.
\end{equation}
Here $\delta_D$ is a derived bound on the circuit implementation error.

\textbf{Step 4. Supply an eigenvalue-threshold operator.}
Let $q_{\mathrm{th}}\in\R\sbra{x}$ be the even threshold polynomial
constructed in the proof of \cite[Lemma II.19]{WGL+24}, with threshold
$\theta$ and accuracy $\eps_{\mathrm{th}}$. Define the positive semidefinite
reference operator
\begin{equation*}
    \eta\coloneqq\widehat\mu q_{\mathrm{th}}\rbra{\widehat\mu}^2.
\end{equation*}
Each algorithm supplies an $n$-qubit subnormalized density operator
$\widetilde\eta$ such that
$\Abs{\widetilde\eta-\eta}\leq\delta_\eta$,
\begin{equation}\label{eq:eigen_threshold_operator}
    \rbra*{\frac{\theta}{4}\rbra*{1-2\eps_{\mathrm{th}}}-\theta^{1/2}\eps_{\mathrm{th}}}
    \Pi_{\supp_{2\theta}\rbra{\widehat\mu}}
    \leq \eta \leq
    \rbra*{\frac{\theta}{4}+\eps_{\mathrm{th}}^2+\theta^{1/2}\eps_{\mathrm{th}}}
    \Pi_{\supp\rbra{\widehat\mu}},
\end{equation}
and
\begin{equation}\label{eq:eta_rank_preservation}
    \max\cbra*{\rank\rbra*{\widetilde{\eta}},\rank\rbra{\eta}}
    \leq\rank\rbra{\widehat\mu}
    \leq\rank\rbra{\rho}+\rank\rbra{\sigma}
    \leq2r.
\end{equation}
The query algorithm prepares $\widetilde\eta$ with
\cref{lem:eigen_threshold_projector}, and the sample algorithm obtains it
from the filter in \cref{lem:eigen_threshold_filter}. Both algorithms
estimate
$y\coloneqq c\tr\rbra*{\widetilde D\widetilde\eta\widetilde D^\dagger}$.
In both branches, $D$ approximates $\nu^{1/(2p)}/8$, and
\cref{eq:core_error} shows that $2^{8+3/p}\theta^{-1}y$ approximates
$\power{\rho}{\sigma}{p}$.

\subsection{Error analysis}\label{sec:error_analysis}

In the proofs, we use the abbreviations introduced in
\cref{sec:construct_unitary}.

We first bound the errors from the outer and inner polynomial approximations
and from the eigenvalue threshold. We then account for circuit
implementation errors and combine these bounds into a single error bound
for $y$, which both algorithms use.

\begin{proposition}\label{prop:p2_approx}
    Fix a constant $p\in\interval[open]{0}{1}$ with $1/p\notin2\Z$
    and a scale $c\in\cbra{1,2}$.
    Let $r$ be a positive integer, and let $\rho$ and $\sigma$ be density operators on the same
    finite-dimensional Hilbert space, each of rank at most $r$.
    For $\eps_{\mathrm{in}},\eps_{\mathrm{out}}\in\interval[open]{0}{1/2}$ and
    $\theta,\eps_{\mathrm{th}}\in\interval[open left]{0}{1/10}$ satisfying
    $32\eps_{\mathrm{th}}^2\leq\theta$, we have
    \begin{equation*}
        \begin{aligned}
            &\Biggl|\tr\rbra*{
                q_{\mathrm{out}}\rbra*{\frac{q_{\mathrm{in}}\rbra{\rho/c}+q_{\mathrm{in}}\rbra{\sigma/c}}{4}}^2
                \frac{\rho+\sigma}{2c}
                q_{\mathrm{th}}\rbra*{\frac{\rho+\sigma}{2c}}^2}\\
            &\qquad-\frac{1}{16}\tr\rbra*{
                \rbra*{\frac{q_{\mathrm{in}}\rbra{\rho/c}+q_{\mathrm{in}}\rbra{\sigma/c}}{4}}^{1/p}
                \frac{\rho+\sigma}{2c}
                q_{\mathrm{th}}\rbra*{\frac{\rho+\sigma}{2c}}^2}\Biggr|\\
            &\qquad\leq O\rbra*{r\eps_{\mathrm{out}}\rbra*{\theta+\eps_{\mathrm{th}}^2+\theta^{1/2}\eps_{\mathrm{th}}}},
        \end{aligned}
    \end{equation*}
    where $q_{\mathrm{in}}$, $q_{\mathrm{out}}$, and $q_{\mathrm{th}}$ are the
    polynomials specified in \cref{sec:construct_unitary}.
\end{proposition}

\begin{proof}
    The spectra of $\widehat\rho$ and $\widehat\sigma$ lie in $\interval{0}{1}$, so
    the bounds on $q_{\mathrm{in}}$ and functional calculus give
    $0\leq\nu\leq\frac{1}{2}I$, so its spectrum lies in
    $\interval{0}{1/2}$. Applying the scalar difference-of-squares
    identity on this spectrum and using the approximation and
    boundedness properties of $q_{\mathrm{out}}$ gives
    \begin{equation}\label{eq:first_part}
        \Abs*{q_{\mathrm{out}}\rbra{\nu}^2-\frac{1}{16}\nu^{1/p}}
        \leq\frac{5}{4}\eps_{\mathrm{out}}.
    \end{equation}
    Since $\eta$ is positive semidefinite, its upper bound in
    \cref{eq:eigen_threshold_operator} and
    $\rank\rbra{\widehat\mu}\leq\rank\rbra{\widehat\rho}+\rank\rbra{\widehat\sigma}\leq2r$
    yield
    \begin{equation}\label{eq:eta_trace_norm_bound}
        \Abs*{\eta}_1=\tr\rbra{\eta}
        \leq2r\rbra*{\frac{\theta}{4}+\eps_{\mathrm{th}}^2
        +\theta^{1/2}\eps_{\mathrm{th}}}.
    \end{equation}
    By cyclicity of the trace, the error in the statement is the absolute
    value of $\tr\rbra{\rbra{q_{\mathrm{out}}\rbra{\nu}^2
    -\nu^{1/p}/16}\eta}$. The operator--trace norm H\"older inequality
    (\cref{fact:holder}) bounds this by
    $\Abs{q_{\mathrm{out}}\rbra{\nu}^2-\nu^{1/p}/16}\Abs{\eta}_1$.
    Combining \cref{eq:first_part,eq:eta_trace_norm_bound} proves the claim.
\end{proof}

\begin{proposition}\label{prop:p1_approx}
    Fix a constant $p\in\interval[open]{0}{1}$ and a scale $c\in\cbra{1,2}$.
    Let $r$ be a positive integer, and let $\rho$ and $\sigma$ be density operators on the same
    finite-dimensional Hilbert space, each of rank at most $r$.
    For $\eps_{\mathrm{in}}\in\interval[open]{0}{1/2}$ and
    $\theta,\eps_{\mathrm{th}}\in\interval[open left]{0}{1/10}$ satisfying
    $32\eps_{\mathrm{th}}^2\leq\theta$, we have
    \begin{equation*}
        \begin{aligned}
            &\Biggl|\tr\rbra*{
                \rbra*{\frac{q_{\mathrm{in}}\rbra{\rho/c}+q_{\mathrm{in}}\rbra{\sigma/c}}{4}}^{1/p}
                \frac{\rho+\sigma}{2c}
                q_{\mathrm{th}}\rbra*{\frac{\rho+\sigma}{2c}}^2}\\
            &\qquad-\tr\rbra*{
                \rbra*{\frac{\rho^p+\sigma^p}{16c^p}}^{1/p}
                \frac{\rho+\sigma}{2c}
                q_{\mathrm{th}}\rbra*{\frac{\rho+\sigma}{2c}}^2}\Biggr|\\
            &\qquad\leq O\rbra*{r^2\eps_{\mathrm{in}}\rbra*{\theta+\eps_{\mathrm{th}}^2+\theta^{1/2}\eps_{\mathrm{th}}}},
        \end{aligned}
    \end{equation*}
    where $q_{\mathrm{in}}$ and $q_{\mathrm{th}}$ are the polynomials
    specified in \cref{sec:construct_unitary}.
\end{proposition}

\begin{proof}
    Positivity of $\widehat\rho$ and $\widehat\sigma$
    gives $\ker\rbra{\widehat\mu}=\ker\rbra{\widehat\rho}\cap\ker\rbra{\widehat\sigma}$
    and $\supp\rbra{\widehat\mu}=\ran\rbra{\widehat\rho}+\ran\rbra{\widehat\sigma}$. Hence
    $\dim\rbra*{\supp\rbra{\widehat\mu}}=\rank\rbra{\widehat\mu} \leq\rank\rbra{\widehat\rho}+\rank\rbra{\widehat\sigma}\leq2r$.
    The subspace $\supp\rbra{\widehat\mu}$ reduces $\widehat\rho$ and $\widehat\sigma$, and
    hence also $\nu$ and $\tau$ by functional calculus. The upper bound
    in \cref{eq:eigen_threshold_operator} implies
    $\supp\rbra{\eta}\subseteq\supp\rbra{\widehat\mu}$. Restricting all
    operators to $\supp\rbra{\widehat\mu}$ therefore leaves both traces in
    the statement unchanged. For the rest of this proof, all operators
    and traces are restricted to this subspace, with the same notation.

    The upper bound on $\eta$ gives
    $\Abs{\eta}_1=\tr\rbra{\eta} \leq2r\rbra{\frac{\theta}{4}+\eps_{\mathrm{th}}^2+ \theta^{1/2}\eps_{\mathrm{th}}}$.
    The bounds on $q_{\mathrm{in}}$, the spectral bounds $0\leq\widehat\rho,\widehat\sigma\leq I$, and
    functional calculus give $\nu,\tau\geq0$,
    $\Abs{\nu}\leq1/2$, and $\Abs{\tau}\leq1/8$.
    Cyclicity of the trace expresses the error in the statement as
    $\abs{\tr\rbra{\rbra{\nu^{1/p}-\tau^{1/p}}\eta}}$.
    The operator--trace norm
    H\"older inequality (\cref{fact:holder}) bounds this by
    $\Abs{\nu^{1/p}-\tau^{1/p}}\Abs{\eta}_1$.

    The approximation property of $q_{\mathrm{in}}$ and the triangle inequality give
    $\Abs{\nu-\tau}\leq\eps_{\mathrm{in}}/2$.
    Since $\supp\rbra{\widehat\mu}$ has dimension at most $2r$, the restricted
    operators $\nu$ and $\tau$ have rank at most $2r$, including when
    $q_{\mathrm{in}}\rbra{0}\ne0$. Moreover,
    $\Abs{\nu}+\Abs{\tau}\leq5/8<1$.
    Since $1/p-1>0$,
    the rank bounds above and \cref{lem:norm-bound} with $\alpha=1/p$ yield
    \begin{equation}\label{eq:another_first_half}
        \Abs*{\nu^{1/p}-\tau^{1/p}} \leq\frac{4r}{p}
        \rbra*{\Abs*{\nu}+\Abs*{\tau}}^{1/p-1}
        \Abs*{\nu-\tau}
        \leq\frac{2r}{p}\eps_{\mathrm{in}}.
    \end{equation}
    Combining \cref{eq:another_first_half} with the trace-norm bound on
    $\eta$ proves the claim.
\end{proof}

\begin{proposition}\label{prop:eigen_threshold_operator}
    Fix a constant $p\in\interval[open]{0}{1}$ and a scale $c\in\cbra{1,2}$.
    Let $r$ be a positive integer, and let $\rho$ and $\sigma$ be density operators on the same
    finite-dimensional Hilbert space, each of rank at most $r$.
    For
    $\theta,\eps_{\mathrm{th}}\in\interval[open left]{0}{1/10}$ satisfying
    $32\eps_{\mathrm{th}}^2\leq\theta$, we have
    \begin{equation*}
        \begin{aligned}
            \Biggl|\frac{4}{\theta}\tr\rbra*{
                \rbra*{\frac{\rho^p+\sigma^p}{16c^p}}^{1/p}
                \frac{\rho+\sigma}{2c}
                q_{\mathrm{th}}\rbra*{\frac{\rho+\sigma}{2c}}^2}
                -\frac{\power{\rho}{\sigma}{p}}{2^{3/p}c}\Biggr|\leq O\rbra*{\eps_{\mathrm{th}}\theta^{-1/2}+r\theta^p},
        \end{aligned}
    \end{equation*}
    where $q_{\mathrm{th}}$ is the polynomial specified in
    \cref{sec:construct_unitary}.
\end{proposition}

\begin{proof}
    Positive powers preserve kernels, so positivity of $\widehat\rho$ and
    $\widehat\sigma$ gives
    $\ker\rbra{\tau}=\ker\rbra{\widehat\rho}\cap\ker\rbra{\widehat\sigma}
    =\ker\rbra{\widehat\mu}$. Thus $\supp\rbra{\tau}=\supp\rbra{\widehat\mu}$ and
    \begin{equation*}
        \tr\rbra*{\tau^{1/(2p)}\Pi_{\supp\rbra{\widehat\mu}}
        \tau^{1/(2p)}}=\tr\rbra*{\tau^{1/p}}
        =\rbra*{\frac{1}{8}}^{1/p}\power{\widehat\rho}{\widehat\sigma}{p}
        =\frac{\power{\rho}{\sigma}{p}}{2^{3/p}c}.
    \end{equation*}
    Since $p\in\interval[open]{0}{1}$, the function
    $x\mapsto\abs{x}^{1/p}$ is continuous and convex on $\R$.
    Applying \cref{thm:convexity_operator_function} to $\widehat\rho^p$ and
    $\widehat\sigma^p$ at $t=1/2$ gives
    \begin{equation*}
        \power{\widehat\rho}{\widehat\sigma}{p} =\tr\rbra*{\rbra*{\frac{\widehat\rho^p+\widehat\sigma^p}{2}}^{1/p}}
        \leq\frac{1}{2}\tr\rbra{\widehat\rho} +\frac{1}{2}\tr\rbra{\widehat\sigma}\leq1.
    \end{equation*}

    By positivity and cyclicity of the trace, the upper bound on $\eta$
    gives
    \begin{equation}\label{eq:second_term}
        4\theta^{-1}\tr\rbra*{\tau^{1/p}\eta}
        -\tr\rbra*{\tau^{1/p}}
        \leq\rbra*{4\eps_{\mathrm{th}}^2\theta^{-1}+4\eps_{\mathrm{th}}\theta^{-1/2}}
        \rbra*{\frac{1}{8}}^{1/p}\power{\widehat\rho}{\widehat\sigma}{p}.
    \end{equation}
    The lower bound on $\eta$ and
    $0\leq\Pi_{\supp_{2\theta}\rbra{\widehat\mu}}\leq I$ similarly give
    \begin{equation}\label{eq:first_term_1}
        \tr\rbra*{\tau^{1/p}}-4\theta^{-1}\tr\rbra*{\tau^{1/p}\eta}
        \leq\rbra*{2\eps_{\mathrm{th}}+4\eps_{\mathrm{th}}\theta^{-1/2}}
        \rbra*{\frac{1}{8}}^{1/p}\power{\widehat\rho}{\widehat\sigma}{p}
        +\tr\rbra*{\tau^{1/p}\rbra*{I-\Pi_{\supp_{2\theta}\rbra{\widehat\mu}}}}.
    \end{equation}
    The spectra of $\widehat\rho$ and $\widehat\sigma$ lie in $\interval{0}{1}$, so
    $\Abs{\tau}\leq1/8$. Functional calculus and
    \cref{lem:density_operator_cdj} therefore give
    \begin{equation*}
        \tau^{1/p}\leq\tau
        =\frac{1}{8}\rbra*{\frac{\widehat\rho^p+\widehat\sigma^p}{2}}
        \leq\frac{1}{8}\widehat\mu^p.
    \end{equation*}
    Let $\widehat\mu=\sum_j\lambda_j\ketbra{\psi_j}{\psi_j}$ be a spectral
    decomposition. Since $\supp\rbra{\tau}=\supp\rbra{\widehat\mu}$, only
    eigenvalues $0<\lambda_j\leq2\theta$ contribute to the remaining
    trace, and
    \begin{equation}\label{eq:first_term_2}
        \tr\rbra*{\tau^{1/p}\rbra*{I-\Pi_{\supp_{2\theta}\rbra{\widehat\mu}}}}
        =\sum_{0<\lambda_j\leq2\theta}
        \bra{\psi_j}\tau^{1/p}\ket{\psi_j}
        \leq\frac{1}{8}\sum_{0<\lambda_j\leq2\theta}\lambda_j^p
        \leq\frac{1}{4}\rbra*{2\theta}^p r,
    \end{equation}
    where we used
    $\rank\rbra{\widehat\mu}\leq\rank\rbra{\widehat\rho}+\rank\rbra{\widehat\sigma}\leq2r$.
    Combining \cref{eq:second_term,eq:first_term_1,eq:first_term_2} yields
    \begin{equation}\label{eq:bound_with_p_power_means}
        \abs*{4\theta^{-1}\tr\rbra*{\tau^{1/p}\eta}
        -\tr\rbra*{\tau^{1/p}}}
        \leq O\rbra*{\eps_{\mathrm{th}}\theta^{-1/2}\power{\widehat\rho}{\widehat\sigma}{p}
        +r\theta^p},
    \end{equation}
    using $32\eps_{\mathrm{th}}^2\leq\theta$ and $\theta\leq1/10$.
    By cyclicity of the trace, the left-hand side is the error in the
    statement. The bound $\power{\widehat\rho}{\widehat\sigma}{p}\leq1$ reduces
    \cref{eq:bound_with_p_power_means} to the claimed estimate.
\end{proof}

We next account for both the nonzero block-encoding errors and the
discrepancy between the threshold operator $\widetilde{\eta}$ and
its reference operator $\eta$. Return to the setting and parameter
assumptions of \cref{sec:construct_unitary}. By
\cref{eq:eta_rank_preservation} and
$\Abs{\widetilde{\eta}-\eta}\leq\delta_\eta$,
we have $\Abs{\widetilde{\eta}-\eta}_1\leq4r\delta_\eta$.
Moreover, \cref{eq:eigen_threshold_operator} and
$32\eps_{\mathrm{th}}^2\leq\theta$ imply
$\tr\rbra{\eta}=O\rbra{r\theta}$. Since $\delta_\eta\leq\theta$, it
follows that $\tr\rbra{\widetilde\eta}\leq\tr\rbra{\eta}+4r\delta_\eta
=O\rbra{r\theta}$. Inserting
$\tr\rbra{D\widetilde{\eta}D^\dagger}$ between the two trace terms and
using cyclicity of the trace, H\"older's inequality in \cref{fact:holder},
and \cref{eq:outer_block_error}, we obtain
\begin{equation}\label{eq:circuit_implementation_error}
    \begin{aligned}
        \abs*{\tr\rbra*{\widetilde D\widetilde{\eta}\widetilde D^\dagger}
        -\tr\rbra*{D\eta D^\dagger}}
        &\leq
        \Abs*{\widetilde D^\dagger\widetilde D-D^\dagger D}
        \tr\rbra{\widetilde{\eta}}
        +\Abs*{D}^2\Abs*{\widetilde{\eta}-\eta}_1\\
        &\leq
        \rbra*{\Abs*{\widetilde D}+\Abs*{D}}\delta_D
        \tr\rbra{\widetilde{\eta}}
        +\Abs*{D}^2\Abs*{\widetilde{\eta}-\eta}_1\\
        &=O\rbra*{r\theta\delta_D+r\delta_\eta}.
    \end{aligned}
\end{equation}

\paragraph{Combined error.}
Combining these bounds gives
    \begin{equation}\label{eq:core_error}
        \abs*{2^{8+3/p}\theta^{-1}y-\power{\rho}{\sigma}{p}}
        \leq O\rbra*{r^2\eps_{\mathrm{in}}+\eps_{\mathrm{th}}\theta^{-1/2}+r\theta^p
        +r\delta_D+r\delta_\eta\theta^{-1}+r\eps_{\mathrm{out}}}.
    \end{equation}
    To prove this bound, set $P\coloneqq\Pi_{\supp\rbra{\widehat\mu}}$.
    By homogeneity of the power mean,
    $\power{\rho}{\sigma}{p}=c\,\power{\widehat\rho}{\widehat\sigma}{p}$.
    By the proof of \cref{prop:eigen_threshold_operator},
    $\power{\widehat\rho}{\widehat\sigma}{p}=2^{3/p}\tr\rbra{\tau^{1/p}}
    =2^{3/p}\tr\rbra{\tau^{1/p}P}$. Since $c\leq2$, the definition of $y$
    and the triangle inequality give
    \begin{align*}
        \abs*{2^{8+3/p}\theta^{-1}y-\power{\rho}{\sigma}{p}}
        \leq2^{9+3/p}\theta^{-1}
        \abs*{\tr\rbra*{\widetilde D\widetilde\eta\widetilde D^\dagger}
        -\tr\rbra*{D\eta D^\dagger}}
        +2\abs*{2^{8+3/p}\theta^{-1}\tr\rbra*{D\eta D^\dagger}
        -2^{3/p}\tr\rbra*{\tau^{1/p}P}}.
    \end{align*}
    By \cref{eq:circuit_implementation_error}, the first term is
    $O\rbra{r\delta_D+r\delta_\eta\theta^{-1}}$.

    Throughout, the condition $32\eps_{\mathrm{th}}^2\leq\theta$ gives
    $\theta+\eps_{\mathrm{th}}^2+\theta^{1/2}\eps_{\mathrm{th}}=O\rbra{\theta}$.

    Since $D=q_{\mathrm{out}}\rbra{\nu}/2$, cyclicity of the trace and
    \cref{prop:p2_approx} bound the outer-polynomial error by
    \begin{equation*}
        \abs*{\tr\rbra*{D\eta D^\dagger}
        -\frac{1}{64}\tr\rbra*{\nu^{1/p}\eta}}
        \leq O\rbra{r\eps_{\mathrm{out}}\theta}
    \end{equation*}
    in the approximate branch. In the exact branch this error is zero,
    so the same upper bound holds. The second absolute value in the
    initial decomposition is therefore at most
    \begin{equation*}
        O\rbra{r\eps_{\mathrm{out}}}
        +2^{2+3/p}\theta^{-1}
        \abs*{\tr\rbra*{\rbra*{\nu^{1/p}-\tau^{1/p}}\eta}}
        +2^{3/p}
        \abs*{\tr\rbra*{\tau^{1/p}\rbra*{4\theta^{-1}\eta-P}}}.
    \end{equation*}
    By cyclicity of the trace, \cref{prop:p1_approx} bounds the middle
    absolute value by $O\rbra{r^2\eps_{\mathrm{in}}\theta}$, and
    \cref{prop:eigen_threshold_operator} bounds the last by
    $O\rbra{\eps_{\mathrm{th}}\theta^{-1/2}+r\theta^p}$.
    Combining these estimates with the implementation error proves
    \cref{eq:core_error}.

\subsection{Query complexity upper bound}\label{sec:query_algorithm}

\begin{theorem}[Query upper bound for the divergence with $p$-power means]\label{thm:query_estimate_power_divergence}
    Fix a polynomial-time computable constant
    $p\in\interval[open]{0}{1}$. Let $\rho$ and $\sigma$ be $n$-qubit
    density operators with a known common integer rank upper bound $r$, and let
    $\eps\in\interval[open]{0}{1}$.
    Given purified quantum query access oracles $\calO_\rho$ and
    $\calO_\sigma$, there is a quantum algorithm that outputs a
    classical estimate of $\dpower{\rho}{\sigma}{p}$ within additive error
    $\eps$ with probability at least $2/3$.

    \begin{samepage}
    \setlength{\emergencystretch}{2em}
    It uses
    $\widetilde{O}\rbra{r^{3/p}\eps^{-(1+2/p)}}$ queries when
    $1/p\in2\Z$ and
    $\widetilde{O}\rbra{r^{2p+3/p}\eps^{-(2p+1+2/p)}}$ queries
    when \mbox{$1/p\notin2\Z$},
    counted in total over $\calO_\rho$ and $\calO_\sigma$.
    \par
    \end{samepage}
\end{theorem}

We now implement the shared construction with $c=1$ in the purified query
model.

\paragraph{Implementation.}
Let $\rho$ and $\sigma$ be $n$-qubit density operators, and let $r$ be a
known upper bound on their ranks.
Suppose $\calO_\rho$ and $\calO_\sigma$ provide purified quantum query
access to $\rho$ and $\sigma$, respectively, and act on $n+a_\rho$ and
$n+a_\sigma$ qubits.
Set $a_0\coloneqq\max\cbra{a_\rho,a_\sigma}$ and $a\coloneqq a_0+1$.
We pad both oracles with idle ancillas so that they act on $n+a_0$ qubits
and reserve one additional qubit as a selector. When constructing the
individual block-encodings, this selector is idle. When preparing the
equal mixture, it controls which oracle is applied. Thus the
state-preparation inputs used in Steps~Q1 and~Q2 share a common
$\rbra{n+a}$-qubit register layout.

\textbf{Step Q1. Construct the block-encodings of $\rho$ and $\sigma$.}
By \cref{lem:block-encode-subnormalized-density-operator}, the padded
oracles yield unitary $\rbra{1,n+a,0}$-block-encodings $U_\rho$ and
$U_\sigma$ of $\rho$ and $\sigma$, respectively. Apply Steps~1--3 of
\cref{sec:construct_unitary} with $c=1$, $m=n+a$,
$U_{\widehat\rho}=U_\rho$, and $U_{\widehat\sigma}=U_\sigma$. This gives
the unitary $U_D$, which acts on $2n+a+5$ qubits.

\textbf{Step Q2. Prepare the average state.}
Let $s$ denote the reserved selector qubit and define the
$\rbra{n+a}$-qubit unitary
\begin{equation*}
    U_\mu\coloneqq
    \rbra*{
        \ketbra{0}{0}_s\otimes\calO_\rho
        +\ketbra{1}{1}_s\otimes\calO_\sigma}
    \rbra*{H_s\otimes I_{n+a_0}}.
\end{equation*}
We denote this unitary by
$\mathsf{AverageState}\rbra{\calO_\rho,\calO_\sigma}$. On the all-zero input,
it prepares
\begin{equation*}
    \frac{\ket{0}_s\otimes\calO_\rho\ket{0}_{n+a_0}
    +\ket{1}_s\otimes\calO_\sigma\ket{0}_{n+a_0}}{\sqrt{2}},
\end{equation*}
where both oracles act on the padded registers.
Tracing out the selector and purification registers gives the density
operator $\mu=\rbra{\rho+\sigma}/2$. The construction uses one controlled
query to each input oracle. Its inverse similarly uses one controlled query
to each inverse oracle.

\textbf{Step Q3. Prepare the eigenvalue-threshold operator.}
By \cref{lem:eigen_threshold_projector}, there exists a nonnegative integer
$a_\eta$ satisfying $n+a_\eta=O\rbra{n+a}$ such that the
$\rbra{n+a_\eta}$-qubit unitary
$U_\eta=\mathsf{EigenThreProj}\rbra{U_\mu,\eps_{\mathrm{th}},\theta,\delta_\eta}$
prepares an $n$-qubit subnormalized density operator $\widetilde{\eta}$.
The reference operator $\eta$ defined in Step~4 of
\cref{sec:construct_unitary} satisfies
$\Abs{\widetilde{\eta}-\eta}\leq\delta_\eta$, and
\cref{eq:eigen_threshold_operator} holds with $\widehat\mu=\mu$. The rank guarantee in
\cref{lem:eigen_threshold_projector} also gives
\cref{eq:eta_rank_preservation}. Thus $\widetilde\eta$
satisfies the requirements of Step~4 of \cref{sec:construct_unitary}.

\textbf{Step Q4. Evolve the threshold operator.}
Since $\widetilde D$ is the actual top-left block of $U_D$, the latter
is an exact $\rbra{1,n+a+5,0}$-block-encoding of $\widetilde D$.
Apply \cref{lem:evolution_density_operator} to $U_\eta$ and $U_D$.
The resulting unitary
$U_X\coloneqq\mathsf{SubDensityOpEvolution}\rbra{U_\eta,U_D}$ acts on
$2n+a+a_\eta+5$ qubits and prepares the subnormalized density operator
$\widetilde D\widetilde\eta\widetilde D^\dagger$ using one call each
to $U_\eta$ and $U_D$. Since $c=1$, its trace is $y$.

\textbf{Step Q5. Estimate the trace.}
Apply \cref{lem:trace_est_subdensity-operator} with accuracy $\eps_{\mathrm{est}}$ and
trace upper bound $1$, which is valid because the prepared operator is
subnormalized. With probability at least $8/\pi^2$, the output
$X\in\R$ of $\mathsf{EstTrSubDensityOp}\rbra{U_X,1,\eps_{\mathrm{est}}}$ satisfies
$\abs{X-y}\leq\eps_{\mathrm{est}}$. This step uses
$O\rbra{\eps_{\mathrm{est}}^{-1}+\eps_{\mathrm{est}}^{-1/2}}$ queries in total to $U_X$ and
$U_X^\dagger$.

\paragraph{The query algorithm.}
For a fixed polynomial-time computable $p\in\interval[open]{0}{1}$,
\cref{alg:p_power_mean} estimates the $p$-power mean
$\power{\rho}{\sigma}{p}$. The branch determined by $p$ is fixed as part
of the algorithm description.

\begin{algorithm}[!ht]
    \caption{$\mathsf{QueryPowerMean}_p\rbra{\calO_\rho, \calO_\sigma, r, \eps_{\mathrm{alg}}}$}
    \label{alg:p_power_mean}
    \small
    \begin{algorithmic}[1]
        \Require Positive integers $n,r$, nonnegative integers
        $a_\rho,a_\sigma$, and an accuracy parameter
        $\eps_{\mathrm{alg}}\in\interval[open]{0}{1}$. The input oracles $\calO_\rho$ and
        $\calO_\sigma$ provide purified quantum query access to
        $n$-qubit density operators $\rho,\sigma$ with known
        rank upper bound $r\geq\max\cbra{\rank\rbra{\rho},\rank\rbra{\sigma}}$
        and act on $n+a_\rho$ and $n+a_\sigma$ qubits, respectively.
        \Ensure A classical estimate of $\power{\rho}{\sigma}{p}$ within
        additive error $O\rbra{\eps_{\mathrm{alg}}}$, with probability at least $8/\pi^2$.

        \State $\eps_{\mathrm{in}}\gets\frac{\eps_{\mathrm{alg}}}{32r^2}$,
        $\eps_{\mathrm{out}}\gets\frac{\eps_{\mathrm{alg}}}{32r}$, and
        $\theta\gets\rbra{\frac{\eps_{\mathrm{alg}}}{32r}}^{1/p}$.
        \State $\eps_{\mathrm{th}}\gets\frac{\eps_{\mathrm{alg}}\sqrt\theta}{8}$,
        $\delta_{\mathrm{out}}\gets\frac{\eps_{\mathrm{alg}}}{128r}$, and
        $\delta_\eta\gets\frac{\eps_{\mathrm{alg}}\theta}{128r}$.
        \State $\eps_{\mathrm{est}}\gets\eps_{\mathrm{alg}}\theta$.
        \State Compute $q_{\mathrm{in}}$ using \cref{lem:modified_pos_approx}
        with $k=1$, $\gamma=p-1$, and accuracy $\eps_{\mathrm{in}}$.
        \If{$\frac{1}{p} \in 2\Z$}
            \State Set $q_{\mathrm{out}}\rbra{x}\gets x^{1/(2p)}/4$.
        \Else
            \State Compute $q_{\mathrm{out}}$ using \cref{lem:modified_pos_approx}
            with $k=\ceil{1/(2p)}$, $\gamma=1/(2p)-k$, and accuracy $\eps_{\mathrm{out}}$.
        \EndIf
        \State $d_{\mathrm{out}}\gets\deg\rbra{q_{\mathrm{out}}}$ and
        $\delta_{\mathrm{in}}\gets\rbra{\frac{\eps_{\mathrm{alg}}}{128rd_{\mathrm{out}}}}^2$.
        \State $a_0\gets\max\cbra{a_\rho,a_\sigma}$ and $a\gets a_0+1$.
        \State Pad both input oracles to $n+a_0$ qubits and reserve the
        additional qubit as the selector for the average-state preparation.
        \State Construct $\rbra{1,n+a,0}$-block-encodings $U_\rho$ and
        $U_\sigma$ using \cref{lem:block-encode-subnormalized-density-operator}.
        \State $V_\rho \gets \mathsf{EigenTrans}\rbra{U_\rho,q_{\mathrm{in}}/2,\delta_{\mathrm{in}}}$
        and $V_\sigma \gets \mathsf{EigenTrans}\rbra{U_\sigma,q_{\mathrm{in}}/2,\delta_{\mathrm{in}}}$.
        \State $U_\nu \gets \mathsf{LCU}\rbra{
        \rbra{H,H},\rbra{V_\rho,V_\sigma}}$.
        \Comment{$U_\nu$ is a $\rbra{2,n+a+3,2\delta_{\mathrm{in}}}$-block-encoding of
        $2\nu$.}
        \State $U_D \gets \mathsf{EigenTrans}\rbra{U_\nu,q_{\mathrm{out}}/2,\delta_{\mathrm{out}}}$.
        \Comment{$U_D$ is a $\rbra{1,n+a+5,\delta_D}$-block-encoding of $D$.}
        \State $U_\mu \gets \mathsf{AverageState}\rbra{\calO_\rho,\calO_\sigma}$.
        \State $U_\eta \gets \mathsf{EigenThreProj}\rbra{U_\mu,\eps_{\mathrm{th}},\theta,\delta_\eta}$.
        \State $U_X \gets \mathsf{SubDensityOpEvolution}\rbra{U_\eta,U_D}$.
        \State $X \gets \mathsf{EstTrSubDensityOp}\rbra{U_X,1,\eps_{\mathrm{est}}}$.
        \State \Return $2^{8+3/p}\theta^{-1}X$.
    \end{algorithmic}
\end{algorithm}

\begin{samepage}
\begin{proposition}[Error of the query algorithm]\label{prop:query_error}
    Fix a polynomial-time computable constant $p\in\interval[open]{0}{1}$.
    Let $r$ be a positive integer, and let $\rho$ and $\sigma$ be $n$-qubit density operators, each of rank
    at most $r$.
    Let $X$ and
    $\eps_{\mathrm{in}},\eps_{\mathrm{out}},\eps_{\mathrm{th}},\theta,
    \delta_{\mathrm{in}},\delta_{\mathrm{out}},\delta_\eta,\eps_{\mathrm{est}}$
    be the estimate and parameters specified in \cref{alg:p_power_mean}.
    If $\abs{X-y}\leq\eps_{\mathrm{est}}$, then
    \begin{align*}
        \abs*{2^{8+3/p}\theta^{-1}X-\power{\rho}{\sigma}{p}}
        \leq O\rbra*{r^2\eps_{\mathrm{in}}+\eps_{\mathrm{th}}\theta^{-1/2}
        +r\theta^p+r\delta_D+r\delta_\eta\theta^{-1}+r\eps_{\mathrm{out}}
        +\eps_{\mathrm{est}}\theta^{-1}},
    \end{align*}
    where $y$ is the trace of the subnormalized density operator prepared
    by $U_X$, $d_{\mathrm{out}}=\deg\rbra{q_{\mathrm{out}}}$, and
    $\delta_D=4d_{\mathrm{out}}\sqrt{\delta_{\mathrm{in}}}+\delta_{\mathrm{out}}$.
\end{proposition}
\end{samepage}

\begin{proof}
    By Step~Q4, $y=\tr\rbra{\widetilde D\widetilde\eta\widetilde D^\dagger}$.
    The triangle inequality gives
    \begin{equation*}
        \abs*{2^{8+3/p}\theta^{-1}X-\power{\rho}{\sigma}{p}}
        \leq2^{8+3/p}\theta^{-1}\abs{X-y}
        +\abs*{2^{8+3/p}\theta^{-1}y-\power{\rho}{\sigma}{p}}.
    \end{equation*}
    Apply \cref{eq:core_error} with $c=1$ to the second term and use
    $\abs{X-y}\leq\eps_{\mathrm{est}}$ for the first.
\end{proof}

\begin{proof}[Proof of \cref{thm:query_estimate_power_divergence}]
    Run \cref{alg:p_power_mean} with internal accuracy
    $\eps_{\mathrm{alg}}\in\interval[open]{0}{1}$.
    In the exact branch, $d_{\mathrm{out}}=1/(2p)$ is a positive integer.
    In the approximate branch, $\eps_{\mathrm{out}}<1/8$ and the target in
    \cref{eq:p2_qsvt_approx} has values $0$ and $1/4$ at the endpoints of
    $\interval{0}{1}$, so $q_{\mathrm{out}}$ cannot be constant. Thus
    $d_{\mathrm{out}}\geq1$ in either branch. The parameter choices satisfy
    $\eps_{\mathrm{in}},\eps_{\mathrm{out}}<1/2$,
    $\theta,\eps_{\mathrm{th}}<1/10$, $32\eps_{\mathrm{th}}^2\leq\theta$,
    $\delta_\eta\leq\theta$, and
    $0<\delta_{\mathrm{in}},\delta_{\mathrm{out}},\delta_\eta<1$,
    as required by \cref{sec:construct_unitary} with $c=1$.

    The choices also give
    $r^2\eps_{\mathrm{in}}=\eps_{\mathrm{alg}}/32$,
    $\eps_{\mathrm{th}}\theta^{-1/2}=\eps_{\mathrm{alg}}/8$,
    $r\theta^p=\eps_{\mathrm{alg}}/32$,
    $r\delta_D=r\rbra{4d_{\mathrm{out}}\sqrt{\delta_{\mathrm{in}}}+\delta_{\mathrm{out}}}=5\eps_{\mathrm{alg}}/128$,
    $r\delta_\eta\theta^{-1}=\eps_{\mathrm{alg}}/128$,
    $r\eps_{\mathrm{out}}=\eps_{\mathrm{alg}}/32$, and
    $\eps_{\mathrm{est}}\theta^{-1}=\eps_{\mathrm{alg}}$.
    Since $0<\eps_{\mathrm{est}}<1$, trace estimation in Step~Q5 gives
    $\abs{X-y}\leq\eps_{\mathrm{est}}$ with probability at least $8/\pi^2>2/3$.
    On this event, \cref{prop:query_error} gives
    $\abs*{2^{8+3/p}\theta^{-1}X-\power{\rho}{\sigma}{p}} \leq O\rbra{\eps_{\mathrm{alg}}}$.
    Trace estimation is the only randomized stage.

    Since $\eps_{\mathrm{est}}<1$, trace estimation uses
    $O\rbra{1/\eps_{\mathrm{est}}}$ queries to $U_X$ and $U_X^\dagger$
    by \cref{lem:trace_est_subdensity-operator}.
    By the construction in \cref{lem:evolution_density_operator},
    each call to $U_X$ (or $U_X^\dagger$) uses one call each to
    $U_\eta$ and $U_D$ (or their inverses).
    The threshold construction uses
    $O\rbra{\theta^{-1}\log\rbra{1/\eps_{\mathrm{th}}}}$ input-oracle
    queries per call to $U_\eta$ by \cref{lem:eigen_threshold_projector}.
    Each call to $U_D$ uses $O\rbra{d_{\mathrm{out}}}$ calls to $U_\nu$,
    each costing $O\rbra{d_{\mathrm{in}}}$ input-oracle queries by
    \cref{thm:qsvt,lem:lcu,lem:block-encode-subnormalized-density-operator}.
    These counts include inverse and controlled calls, giving total query
    complexity
    \begin{equation*}
        O\rbra*{\frac{1}{\eps_{\mathrm{est}}}\rbra*{
        \frac{1}{\theta}\log\rbra*{\frac{1}{\eps_{\mathrm{th}}}}
        +d_{\mathrm{in}}d_{\mathrm{out}}}}.
    \end{equation*}

    The polynomial degree bounds and the choice of $\theta$ give
    \begin{equation*}
        d_{\mathrm{in}}=O\rbra*{\frac{r^{2/p}}{\eps_{\mathrm{alg}}^{1/p}}},\qquad
        \theta^{-1}=O\rbra*{\frac{r^{1/p}}{\eps_{\mathrm{alg}}^{1/p}}}.
    \end{equation*}
    Moreover, $d_{\mathrm{out}}=1/(2p)$ when $1/p\in2\Z$
    and $d_{\mathrm{out}}=O\rbra{r^{2p}\eps_{\mathrm{alg}}^{-2p}}$
    when $1/p\notin2\Z$.
    With $\eps_{\mathrm{est}}^{-1}=O\rbra{r^{1/p}/\eps_{\mathrm{alg}}^{1+1/p}}$,
    the threshold contribution is at most
    $\widetilde{O}\rbra{r^{2/p}/\eps_{\mathrm{alg}}^{1+2/p}}$ and the outer-circuit
    contribution is at most
    $O\rbra{d_{\mathrm{out}}r^{3/p}/\eps_{\mathrm{alg}}^{1+2/p}}$.
    Since $r^{2/p}\leq d_{\mathrm{out}}r^{3/p}$ for
    $r,d_{\mathrm{out}}\geq1$, the threshold contribution is absorbed, giving
    $\widetilde{O}\rbra{\frac{d_{\mathrm{out}}r^{3/p}} {\eps_{\mathrm{alg}}^{1+2/p}}}$.
    Substituting the bound on $d_{\mathrm{out}}$ in each branch gives the corresponding query bounds
    in terms of $\eps_{\mathrm{alg}}$.

    By \cref{lem:modified_pos_approx}, the polynomials can be computed
    in classical polynomial time; in the exact branch,
    $q_{\mathrm{out}}$ is explicit.
    The implementation precisions $\delta_{\mathrm{in}}$,
    $\delta_{\mathrm{out}}$, and $\delta_\eta$ affect only the classical
    circuit descriptions, through factors polynomial in their logarithms
    (\cref{thm:qsvt,lem:eigen_threshold_projector}). For fixed $p$, these
    logarithms are $O\rbra{\log\rbra{r/\eps_{\mathrm{alg}}}}$ and do not
    change the query bounds.

    Choose $\eps_{\mathrm{alg}}$ to be a sufficiently small fixed positive
    rational multiple of $\eps$, depending only on $p$, and subtract the
    algorithm's output from $1$. Since
    $\dpower{\rho}{\sigma}{p}=1-\power{\rho}{\sigma}{p}$, this gives
    additive error at most $\eps$ with the same success probability and
    no additional queries. Substituting this choice of $\eps_{\mathrm{alg}}$
    gives the stated query bounds, with constants depending only on $p$.
\end{proof}

\subsection{Sample complexity upper bound}
\label{sec:sample_algorithm}

\begin{theorem}[Sample upper bound for the divergence with $p$-power means]
\label{thm:sample_estimate_power_divergence}
    Fix a polynomial-time computable constant
    $p\in\interval[open]{0}{1}$. Let $\rho$ and $\sigma$ be $n$-qubit
    density operators with a known common integer rank upper bound $r$, and let
    $\eps\in\interval[open]{0}{1}$.
    Given sample access to $\rho$ and $\sigma$, there is a quantum
    algorithm that outputs a classical estimate of
    $\dpower{\rho}{\sigma}{p}$ within additive error $\eps$ with
    probability at least $2/3$. It uses
    $\widetilde{O}\rbra{r^{6/p}\eps^{-(3+4/p)}}$ copies when
    $1/p\in2\Z$ and
    $\widetilde{O}\rbra{r^{4p+6/p}\eps^{-(4p+3+4/p)}}$ copies when
    $1/p\notin2\Z$,
    counted in total over the two input states. Its quantum gate complexity satisfies the same
    bounds with an additional factor of $n$, and its
    classical circuit-generation time is polynomial in $n,r,1/\eps$
    for fixed $p$.
\end{theorem}

We now implement the shared construction with $c=2$ using only independent
copies of the input states. The resulting sample complexity is independent
of the dimension.
Here the threshold filter and the outer power transformation are both
constructed from formal block-encoding oracles. This gives a
circuit to which \cref{lem:multi-samplizer} applies, while a fresh
sample supplies the input. We also use the small acceptance
probability to bound the number of trials.

We first obtain a threshold filter whose input is an exact
block-encoding. This form follows from the construction underlying
\cref{lem:eigen_threshold_projector}.

\begin{lemma}[Eigenvalue-threshold filter from a block-encoding]
\label{lem:eigen_threshold_filter}
    Fix integers $n\geq1$ and $a\geq0$. Let $U_M$ be a unitary
    $\rbra{1,a,0}$-block-encoding of an $n$-qubit subnormalized density
    operator $M$. Let
    $\delta,\eps\in\interval[open left]{0}{1/10}$ satisfy
    $32\eps^2\leq\delta$, and let
    $\delta^\prime\in\interval[open]{0}{1}$. There is an $\rbra{n+a+2}$-qubit unitary
    $V=\mathsf{EigenThresholdFilter}
    \rbra{U_M,\eps,\delta,\delta^\prime}$
    with all-zero signal block
    $A_1\coloneqq \rbra{\bra{0}_{a+2}\otimes I_n} V \rbra{\ket{0}_{a+2}\otimes I_n}$
    such that $\widetilde\eta\coloneqq A_1MA_1^\dagger$ satisfies
    $\Abs{\widetilde\eta-\eta}\leq\delta^\prime$ for a positive
    semidefinite operator $\eta$ with
    \begin{equation*}
        \rbra*{\frac{\delta}{4}\rbra*{1-2\eps}
        -\delta^{1/2}\eps}
        \Pi_{\supp_{2\delta}\rbra{M}}
        \leq\eta\leq
        \rbra*{\frac{\delta}{4}+\eps^2
        +\delta^{1/2}\eps}
        \Pi_{\supp\rbra{M}}.
    \end{equation*}
    Moreover,
    $\max\cbra{\rank\rbra{\widetilde\eta},
    \rank\rbra{\eta}}\leq\rank\rbra{M}$.
    The filter uses $O\rbra{d}$ queries in total to $U_M$ and
    $U_M^\dagger$, one additional query to controlled-$U_M$, and
    $O\rbra{\rbra{a+1}d}$ one- and two-qubit gates, where
    $d=O\rbra{\delta^{-1}\log\rbra{1/\eps}}$. A description of $V$
    can be computed classically in time
    $\poly\rbra{a,d,\log\rbra{1/\delta^\prime}}$.
\end{lemma}

\begin{proof}
    In the proof of \cite[Lemma II.19]{WGL+24}, the preparation oracle is
    first converted into an exact block-encoding of the input operator.
    Every subsequent polynomial and QSVT step depends on that oracle only
    through this block-encoding, so starting from the supplied $U_M$ skips
    the initial conversion. The cited proof constructs an even polynomial
    $q_{\mathrm{th}}\in\R\sbra{x}$ of degree $d$ with norm at most one on
    $\interval{-1}{1}$. This is the polynomial specified in Step~4 of
    \cref{sec:construct_unitary}, with $\theta=\delta$ and
    $\eps_{\mathrm{th}}=\eps$. Set $\eta\coloneqq Mq_{\mathrm{th}}\rbra{M}^2$.
    Applying
    the parity-refined QSVT theorem used there
    \cite[Theorem~II.3]{WGL+24} to $U_M$ with accuracy
    $\delta^\prime/2$ gives $V$ with signal block $A_1$ satisfying
    $\Abs{A_1-q_{\mathrm{th}}\rbra{M}}\leq\delta^\prime/2$.
    Since $M$, $A_1$, and
    $q_{\mathrm{th}}\rbra{M}$ have norm at most one,
    $\Abs{\widetilde\eta-\eta}\leq2\Abs{A_1-q_{\mathrm{th}}\rbra{M}}
    \leq\delta^\prime$.
    The scalar functional-calculus argument in the cited proof, which uses
    only $M\geq0$ and $\operatorname{spec}\rbra{M}\subseteq
    \interval{0}{1}$, gives the stated sandwich bounds for
    $\eta=Mq_{\mathrm{th}}\rbra{M}^2$. The rank bounds follow because matrix
    multiplication cannot increase rank. The query, gate, and description bounds
    follow from the same QSVT construction.
\end{proof}

\paragraph{Implementation.}
The following steps describe one trial of the sample algorithm.

\textbf{Step S1. Fix formal block-encodings of $\rho/2$ and $\sigma/2$.}
Let $U_{\widehat\rho}$ and $U_{\widehat\sigma}$ be any pair of exact
unitary $\rbra{1,4,0}$-block-encodings of $\widehat\rho=\rho/2$ and
$\widehat\sigma=\sigma/2$. The width $m=4$ matches the oracle width in
\cref{lem:multi-samplizer}, and the construction and error bounds hold for
every such pair. Apply Steps~1--3 of \cref{sec:construct_unitary} with
$c=2$ and $m=4$. This gives the unitary $U_D$.

\textbf{Step S2. Construct the threshold filter.}
By \cref{lem:lcu}, an LCU construction with coefficients
$\rbra{1/2,1/2}$ gives $U_{\widehat\mu}$. This unitary is an exact
normalization-$1$ block-encoding of $\widehat\mu=\rbra{\widehat\rho+\widehat\sigma}/2$.
Apply \cref{lem:eigen_threshold_filter} to $U_{\widehat\mu}$ with parameters
$\eps=\eps_{\mathrm{th}}$, $\delta=\theta$, and
$\delta^\prime=\delta_\eta$, and denote the resulting filter by $V$ and
its top-left block by $A_1$. Let
$\widetilde\eta\coloneqq A_1\widehat\mu A_1^\dagger$, and let $\eta$ be the reference
operator supplied by the lemma. They satisfy
$\Abs{\widetilde\eta-\eta}\leq\delta_\eta$,
\cref{eq:eigen_threshold_operator}, and \cref{eq:eta_rank_preservation}.
Thus $\widetilde\eta$ satisfies the requirements of Step~4 of
\cref{sec:construct_unitary}.

\textbf{Step S3. Form a Bernoulli experiment.}
Let $\calC^{U_{\widehat\rho},U_{\widehat\sigma}}$ be the unitary
circuit that applies $V$ and then $U_D$, using disjoint ancillary
registers. Supply it with all-zero ancillas and an input $\mu=2\widehat\mu$
prepared by choosing a fresh copy of $\rho$ or $\sigma$ uniformly at random.
After the circuit, accept exactly when every signal
ancilla is measured as zero. Since $A_1\mu A_1^\dagger=2\widetilde\eta$,
the acceptance probability is
$\tr\rbra{\widetilde DA_1\mu A_1^\dagger\widetilde D^\dagger}
=2\tr\rbra{\widetilde D\widetilde\eta\widetilde D^\dagger}=y$.

\textbf{Step S4. Apply the multi-samplizer.}
Apply \cref{lem:multi-samplizer} to $\calC$ with error
$\zeta$. Each trial supplies a fresh input $\mu$ independently of the
copies consumed by the samplizer, then performs the same acceptance
measurement. Denote its acceptance probability by $\widetilde y$.
The diamond-norm guarantee in \cref{def:multi-samplizer} holds on an
arbitrary external input, so it gives $\abs{\widetilde y-y}\leq\zeta$.
Every trial is counted, including those in which the signal measurement
fails. The algorithm does not postselect and retry until success.

\paragraph{The sample algorithm.}
For a fixed polynomial-time computable $p\in\interval[open]{0}{1}$,
\cref{alg:sample_power_mean} estimates $\power{\rho}{\sigma}{p}$ from
independent copies of $\rho$ and $\sigma$.
The branch determined by $p$ and the integer $\ceil{1/p}$ are fixed as
part of the algorithm description. The pseudocode uses the
acceptance-probability bound in \cref{lem:sample_acceptance}.

\begin{algorithm}[!ht]
    \caption{$\mathsf{SamplePowerMean}_p\rbra{\rho,\sigma,r,\eps_{\mathrm{alg}}}$}
    \label{alg:sample_power_mean}
    \small
    \begin{algorithmic}[1]
        \Require Positive integers $n,r$ and an accuracy parameter
        $\eps_{\mathrm{alg}}\in\interval[open]{0}{1}$. Independent copies of
        $n$-qubit density operators $\rho,\sigma$ with known
        rank upper bound $r\geq\max\cbra{\rank\rbra{\rho},\rank\rbra{\sigma}}$.
        \Ensure A classical estimate of $\power{\rho}{\sigma}{p}$ within
        additive error $O\rbra{\eps_{\mathrm{alg}}}$, with probability at least $3/4$.

        \State $\eps_{\mathrm{in}}\gets\frac{\eps_{\mathrm{alg}}}{32r^2}$,
        $\eps_{\mathrm{out}}\gets\frac{\eps_{\mathrm{alg}}}{32r}$, and
        $\theta\gets\rbra{\frac{\eps_{\mathrm{alg}}}{32r}}^{1/p}$.
        \State $\eps_{\mathrm{th}}\gets\frac{\eps_{\mathrm{alg}}\sqrt\theta}{8}$,
        $\delta_{\mathrm{out}}\gets\frac{\eps_{\mathrm{alg}}}{128r}$, and
        $\delta_\eta\gets\frac{\eps_{\mathrm{alg}}\theta}{128r}$.
        \State $\eps_{\mathrm{stat}}\gets\frac{\eps_{\mathrm{alg}}\theta}{128}$ and
        $\zeta\gets\frac{\eps_{\mathrm{alg}}\theta}{128}$.
        \State Compute $q_{\mathrm{in}}$ using \cref{lem:modified_pos_approx}
        with $k=1$, $\gamma=p-1$, and accuracy $\eps_{\mathrm{in}}$.
        \If{$\frac{1}{p} \in 2\Z$}
            \State Set $q_{\mathrm{out}}\rbra{x}\gets x^{1/(2p)}/4$.
        \Else
            \State Compute $q_{\mathrm{out}}$ using \cref{lem:modified_pos_approx}
            with $k=\ceil{1/(2p)}$, $\gamma=1/(2p)-k$, and accuracy $\eps_{\mathrm{out}}$.
        \EndIf
        \State $d_{\mathrm{out}}\gets\deg\rbra{q_{\mathrm{out}}}$ and
        $\delta_{\mathrm{in}}\gets\rbra{\frac{\eps_{\mathrm{alg}}}{128rd_{\mathrm{out}}}}^2$.
        \State Let $U_{\widehat\rho}$ and $U_{\widehat\sigma}$ be formal
        $\rbra{1,4,0}$-block-encodings of $\rho/2$ and $\sigma/2$.
        \Comment{Oracle gates of \cref{lem:multi-samplizer}.}
        \State $V_\rho \gets \mathsf{EigenTrans}\rbra{U_{\widehat\rho},q_{\mathrm{in}}/2,\delta_{\mathrm{in}}}$
        and $V_\sigma \gets \mathsf{EigenTrans}\rbra{U_{\widehat\sigma},q_{\mathrm{in}}/2,\delta_{\mathrm{in}}}$.
        \State $U_\nu \gets \mathsf{LCU}\rbra{\rbra{H,H},\rbra{V_\rho,V_\sigma}}$.
        \Comment{$U_\nu$ is a $\rbra{2,7,2\delta_{\mathrm{in}}}$-block-encoding of $2\nu$.}
        \State $U_D \gets \mathsf{EigenTrans}\rbra{U_\nu,q_{\mathrm{out}}/2,\delta_{\mathrm{out}}}$.
        \Comment{$U_D$ is a $\rbra{1,9,\delta_D}$-block-encoding of $D$.}
        \State $U_{\widehat\mu} \gets \mathsf{LCU}\rbra{\rbra{H,H},\rbra{U_{\widehat\rho},U_{\widehat\sigma}}}$.
        \Comment{$U_{\widehat\mu}$ is a $\rbra{1,5,0}$-block-encoding of $\widehat\mu$.}
        \State $V \gets \mathsf{EigenThresholdFilter}\rbra{U_{\widehat\mu},\eps_{\mathrm{th}},\theta,\delta_\eta}$.
        \State Let $\calC^{U_{\widehat\rho},U_{\widehat\sigma}}$ be the unitary
        circuit that applies $V$ and then $U_D$ on disjoint ancillas.
        \State $N\gets\ceil*{\frac{2^{16}\ceil{1/p}}{\eps_{\mathrm{alg}}^2\theta}}$.
        \For{$i=1,\ldots,N$}
            \State Prepare a fresh input $\mu$ by choosing a copy of
            $\rho$ or $\sigma$ uniformly at random.
            \State Apply $\mathsf{Samplize}_{\zeta}
            \langle\calC^{U_{\widehat\rho},U_{\widehat\sigma}}\rangle
            \sbra{\rho,\sigma}$ to this input and all-zero ancillas,
            using fresh independent copies.
            \State $X_i\gets1$ if every signal ancilla is measured as zero,
            and $X_i\gets0$ otherwise.
        \EndFor
        \State $\overline{X}\gets N^{-1}\sum_{i=1}^N X_i$.
        \State \Return $2^{8+3/p}\theta^{-1}\overline{X}$.
    \end{algorithmic}
\end{algorithm}

\begin{lemma}[Acceptance probability]\label{lem:sample_acceptance}
    Let $\rho$ and $\sigma$ be $n$-qubit density operators of rank at
    most $r$, and let $\eps_{\mathrm{alg}}\in\interval[open]{0}{1}$.
    The acceptance probability $\widetilde y$ of one trial of
    \cref{alg:sample_power_mean} satisfies
    $0\leq\widetilde y\leq\frac{\theta}{p}$.
\end{lemma}

\begin{proof}
    Write $\calS=\supp\rbra{\widehat\mu}$ and $P=\Pi_{\calS}$.
    The parameter choices and Step~S2 give
    \begin{equation*}
        0\leq\eta\leq\frac{\theta}{2}P,\qquad
        \Abs*{\widetilde\eta-\eta}_1
        \leq4r\delta_\eta=\frac{\eps_{\mathrm{alg}}\theta}{32},\qquad
        \tr\rbra{\widetilde\eta}\leq2r\theta.
    \end{equation*}
    By \cref{eq:outer_block_error} and the choices in
    \cref{alg:sample_power_mean},
    \begin{equation*}
        \Abs*{\widetilde D-D}
        \leq\frac{5\eps_{\mathrm{alg}}}{128r}\leq\frac{\eps_{\mathrm{alg}}}{16r},
        \qquad \Abs*{D}\leq\frac12,
        \qquad \Abs*{\widetilde D}\leq1.
    \end{equation*}
    Put $y_0=2\tr\rbra{D\eta D^\dagger}$.
    Splitting the difference at $2\tr\rbra{D\widetilde{\eta}D^\dagger}$
    and using cyclicity of the trace and H\"older's inequality in
    \cref{fact:holder} gives
    \begin{equation*}
        \begin{aligned}
            \abs*{y-y_0}
            \leq & 2\rbra*{
            \Abs*{\widetilde D^\dagger\widetilde D-D^\dagger D}
            \tr\rbra{\widetilde{\eta}}
            +\Abs*{D}^2\Abs*{\widetilde{\eta}-\eta}_1}\\
            \leq & 2\rbra*{
            \rbra*{\Abs*{\widetilde D}+\Abs*{D}}
            \Abs*{\widetilde D-D}\tr\rbra{\widetilde{\eta}}
            +\Abs*{D}^2\Abs*{\widetilde{\eta}-\eta}_1}\\
            \leq & 2\rbra*{\frac32\frac{\eps_{\mathrm{alg}}}{16r}\cdot2r\theta
            +\frac14\frac{\eps_{\mathrm{alg}}\theta}{32}}\\
            \leq & \frac{\eps_{\mathrm{alg}}\theta}{2}.
        \end{aligned}
    \end{equation*}

    On $\calS$, the estimate in the proof of \cref{prop:p1_approx} yields
    \begin{equation*}
        \tr\rbra*{\left.\nu^{1/p}\right\vert_{\calS}}
        \leq\tr\rbra{\tau^{1/p}}
        +2r\cdot\frac{2r\eps_{\mathrm{in}}}{p}
        \leq1+\frac{\eps_{\mathrm{alg}}}{8p}.
    \end{equation*}
    Here $\left.\nu^{1/p}\right\vert_{\calS}$ denotes the restriction of
    $\nu^{1/p}$ to $\calS$.
    This requires no rank bound on $\nu$ outside $\calS$.
    The bound in \cref{eq:first_part} applies in both branches, since
    its left-hand side is zero in the exact branch. Together with
    $D=q_{\mathrm{out}}\rbra{\nu}/2$ and $\eta\leq\theta P/2$, it gives
    \begin{equation*}
        y_0\leq\frac{\theta}{4}
        \rbra*{\frac{1+\eps_{\mathrm{alg}}/(8p)}{16}+\frac{5\eps_{\mathrm{alg}}}{64}}
        \leq\frac{\theta}{4}\rbra*{1+\frac{\eps_{\mathrm{alg}}}{8p}}.
    \end{equation*}
    Finally, $\abs{\widetilde y-y}\leq\zeta=
    \eps_{\mathrm{alg}}\theta/128$ implies
    \begin{equation*}
        \widetilde y
        \leq\theta\rbra*{\frac14+\frac{\eps_{\mathrm{alg}}}{32p}
        +\frac{\eps_{\mathrm{alg}}}{2}+\frac{\eps_{\mathrm{alg}}}{128}}
        \leq\frac{\theta}{p},
    \end{equation*}
    since $0<p,\eps_{\mathrm{alg}}<1$. Nonnegativity follows because
    $\widetilde y$ is a probability.
\end{proof}

\begin{proposition}[Error of the sample algorithm]\label{prop:sample_error}
    Fix a polynomial-time computable constant $p\in\interval[open]{0}{1}$.
    Let $r$ be a positive integer, and let $\rho$ and $\sigma$ be $n$-qubit density operators, each of rank
    at most $r$.
    Let $\overline X$ and
    $\eps_{\mathrm{in}},\eps_{\mathrm{out}},\eps_{\mathrm{th}},\theta,
    \delta_{\mathrm{in}},\delta_{\mathrm{out}},\delta_\eta,\eps_{\mathrm{stat}},\zeta$
    be the estimate and parameters specified in \cref{alg:sample_power_mean},
    and let $\widetilde y$ be the actual acceptance probability of one trial.
    If $\abs{\overline X-\widetilde y}\leq\eps_{\mathrm{stat}}$, then
    \begin{align*}
        \abs*{2^{8+3/p}\theta^{-1}\overline X-\power{\rho}{\sigma}{p}}
        &\leq O\bigl(r^2\eps_{\mathrm{in}}+\eps_{\mathrm{th}}\theta^{-1/2}
        +r\theta^p+r\delta_D +r\delta_\eta\theta^{-1}+r\eps_{\mathrm{out}}
        +\rbra{\eps_{\mathrm{stat}}+\zeta}\theta^{-1}\bigr),
    \end{align*}
    where $d_{\mathrm{out}}=\deg\rbra{q_{\mathrm{out}}}$ and
    $\delta_D=4d_{\mathrm{out}}\sqrt{\delta_{\mathrm{in}}}+\delta_{\mathrm{out}}$.
\end{proposition}

\begin{proof}
    By Steps~S3--S4, $y=2\tr\rbra{\widetilde D\widetilde\eta\widetilde D^\dagger}$
    is the acceptance probability before samplization, and
    $\abs{\widetilde y-y}\leq\zeta$. Hence
    $\abs{\overline X-y}\leq\eps_{\mathrm{stat}}+\zeta$.
    The triangle inequality gives
    \begin{equation*}
        \abs*{2^{8+3/p}\theta^{-1}\overline X-\power{\rho}{\sigma}{p}}
        \leq2^{8+3/p}\theta^{-1}\abs{\overline X-y}
        +\abs*{2^{8+3/p}\theta^{-1}y-\power{\rho}{\sigma}{p}}.
    \end{equation*}
    Apply \cref{eq:core_error} with $c=2$ to the second term and use
    $\abs{\overline X-y}\leq\eps_{\mathrm{stat}}+\zeta$ for the first.
\end{proof}

\begin{proof}[Proof of \cref{thm:sample_estimate_power_divergence}]
    Run \cref{alg:sample_power_mean} with internal accuracy
    $\eps_{\mathrm{alg}}\in\interval[open]{0}{1}$.
    Its common parameter choices coincide with those of
    \cref{alg:p_power_mean}. The checks in the proof of
    \cref{thm:query_estimate_power_divergence} therefore also apply with $c=2$:
    the parameter conditions of \cref{sec:construct_unitary} hold,
    $d_{\mathrm{out}}\geq1$, and each of the six construction-error terms
    in \cref{prop:sample_error} is $O\rbra{\eps_{\mathrm{alg}}}$.
    Also, $\eps_{\mathrm{stat}},\zeta\in\interval[open]{0}{1}$ and
    $\rbra{\eps_{\mathrm{stat}}+\zeta}\theta^{-1}=\eps_{\mathrm{alg}}/64$.

    Each trial uses fresh independent
    copies, so $X_1,\ldots,X_N$ are independent Bernoulli variables
    with common mean $\widetilde y$. By
    \cref{lem:sample_acceptance},
    $\Var\rbra{\overline{X}} =\frac{\widetilde y(1-\widetilde y)}{N} \leq\frac{\theta}{pN}$.
    Chebyshev's inequality and the choice of $N$ give
    \begin{equation*}
        \Prob\sbra*{\abs*{\overline{X}-\widetilde y}>\eps_{\mathrm{stat}}}
        \leq\frac{\theta}{pN\eps_{\mathrm{stat}}^2}
        \leq\frac{1}{4p\ceil{1/p}}
        \leq\frac14.
    \end{equation*}
    On the complementary event, \cref{prop:sample_error} gives
    $\abs*{2^{8+3/p}\theta^{-1}\overline{X}-\power{\rho}{\sigma}{p}} \leq O\rbra{\eps_{\mathrm{alg}}}$.

    We next count copies. By
    \cref{thm:qsvt,lem:eigen_threshold_filter}, one trial uses
    $Q=\widetilde{O}\rbra*{\theta^{-1} +d_{\mathrm{in}}d_{\mathrm{out}}}$
    formal queries. With $\zeta=\eps_{\mathrm{alg}}\theta/128$,
    \cref{lem:multi-samplizer} uses
    $\widetilde{O}\rbra{Q^2/(\eps_{\mathrm{alg}}\theta)}$ copies per trial;
    the additional input copy is absorbed by this bound.
    Since $N=O\rbra{1/(\eps_{\mathrm{alg}}^2\theta)}$ for fixed $p$,
    the total sample cost is
    \begin{equation}\label{eq:sample_cost_before_substitution}
        \widetilde{O}\rbra*{
        \frac{\rbra{\theta^{-1}+d_{\mathrm{in}}d_{\mathrm{out}}}^2}
        {\eps_{\mathrm{alg}}^3\theta^2}}.
    \end{equation}
    Reserving enough copies of each state for all choices of the input
    register changes this cost by at most a constant factor.

    The polynomial degree bounds and the choice of $\theta$ give
    \begin{equation*}
        d_{\mathrm{in}}=O\rbra*{\frac{r^{2/p}}{\eps_{\mathrm{alg}}^{1/p}}},\qquad
        \theta^{-1}=O\rbra*{\frac{r^{1/p}}{\eps_{\mathrm{alg}}^{1/p}}}.
    \end{equation*}
    Moreover, $d_{\mathrm{out}}=1/(2p)$ when $1/p\in2\Z$
    and $d_{\mathrm{out}}=O\rbra{r^{2p}\eps_{\mathrm{alg}}^{-2p}}$
    when $1/p\notin2\Z$.
    Since $r,d_{\mathrm{out}}\geq1$, these bounds absorb the threshold
    contribution and give
    $Q=\widetilde{O}\rbra{d_{\mathrm{out}}r^{2/p}\eps_{\mathrm{alg}}^{-1/p}}$.
    Substitution into \cref{eq:sample_cost_before_substitution} gives
    a total sample cost of
    \begin{equation*}
        \widetilde{O}\rbra*{
        \frac{d_{\mathrm{out}}^2r^{6/p}}{\eps_{\mathrm{alg}}^{3+4/p}}}.
    \end{equation*}
    All non-oracle operations in one trial use $O\rbra{nQ}$ gates, since
    the formal oracles have four auxiliary qubits and each QSVT or LCU
    construction adds only a
    constant number of auxiliary qubits. The gate bound in
    \cref{lem:multi-samplizer} therefore gives
    $\widetilde{O}\rbra{nd_{\mathrm{out}}^2r^{6/p}/\eps_{\mathrm{alg}}^{3+4/p}}$
    quantum gates. Substituting the bound on $d_{\mathrm{out}}$ in each
    branch gives the sample and gate bounds
    in terms of $\eps_{\mathrm{alg}}$ for constant success probability. The polynomial
    and QSVT constructions have classical circuit-generation time
    polynomial in $n,r,1/\eps_{\mathrm{alg}}$ for fixed $p$.

    Choose $\eps_{\mathrm{alg}}$ to be a sufficiently small fixed positive
    rational multiple of $\eps$, depending only on $p$, and subtract the
    algorithm's output from $1$. This estimates $\dpower{\rho}{\sigma}{p}$
    within additive error $\eps$ with probability at least $3/4>2/3$.
    Substitution gives the stated resource bounds, with constants depending
    only on $p$. Clipping the estimate to $\interval{0}{1}$ preserves the
    error guarantee.
\end{proof}

\section{Computational hardness}

We prove that low-rank quantum state distinguishability with respect to
the quantum divergence with $p$-power means is $\BQP$-complete under the
threshold conditions in \cref{thm:completeness_lowrank_dpower}.
Hardness already holds for pure states.

\begin{definition}[$\PowerLowQSD_p$]
    Fix a polynomial-time computable constant
    $p\in\interval[open]{0}{1}$. Let $r,a,b$ be efficiently computable
    functions on the positive integers, where $r$ is a known positive
    integer-valued rank bound and the threshold functions $a,b$ take
    values in $\interval{0}{1}$. Assume
    $1\leq r\rbra{n}\leq\poly\rbra{n}$ and $b\rbra{n}<a\rbra{n}$ for
    every positive integer $n$.

    For each positive integer $n$, an instance of
    $\PowerLowQSD_p\sbra{a,b}$ consists of classical descriptions of two
    polynomial-size unitary quantum circuits $Q_\rho$ and $Q_\sigma$ acting
    on $m$ qubits, where $m\geq n$ is polynomially bounded in $n$, and each
    circuit has $n$ designated output qubits. Let $\rho$ and $\sigma$ be
    their reduced output states obtained by applying the circuits to
    $\ket{0}_m$ and tracing out the non-output qubits. The states
    satisfy $\max\cbra{\rank\rbra{\rho},\rank\rbra{\sigma}}\leq r\rbra{n}$.
    The task is to distinguish the following two cases, under the promise
    that one holds:
    \begin{itemize}[nosep,beginpenalty=10000,midpenalty=10000]
    \item \textup{(Yes)} $\dpower{\rho}{\sigma}{p} \geq a\rbra{n}$.
    \item \textup{(No)} $\dpower{\rho}{\sigma}{p} \leq b\rbra{n}$.
    \end{itemize}
\end{definition}

\begin{theorem}\label{thm:completeness_lowrank_dpower}
    Fix a polynomial-time computable constant
    $p\in\interval[open]{0}{1}$, and let $r,a,b$ be efficiently computable
    functions on the positive integers. Assume that $r$ is
    integer-valued with $1\leq r\rbra{n}\leq\poly\rbra{n}$, that $a$ and
    $b$ take values in $\interval{0}{1}$, and that $b\rbra{n}<a\rbra{n}$
    for every positive integer $n$. Suppose further that, for all
    sufficiently large $n$, $a\rbra{n}-b\rbra{n}\geq1/\poly\rbra{n}$ and
    \begin{equation*}
        a\rbra{n}\leq\frac{1/p-1}{2^{1/p+1}}
        \rbra*{1-2^{-n-1}}^2,\qquad
        b\rbra{n}\geq\frac{1}{2p}2^{-2n-2}.
    \end{equation*}
    Then $\PowerLowQSD_p\sbra{a,b}$ with rank bound $r\rbra{n}$ is
    $\BQP$-complete.
\end{theorem}

We prove the hardness and containment parts of
\cref{thm:completeness_lowrank_dpower} in turn.

\subsection{\texorpdfstring{$\BQP$}{BQP}-hardness}

We prove hardness for the pure-state restriction.

\begin{definition}[$\PowerPureQSD_p$]
    The problem $\PowerPureQSD_p\sbra{a,b}$ is the restriction of
    $\PowerLowQSD_p\sbra{a,b}$ to instances whose reduced output states
    are pure. Since pure states have rank one, this restriction does not
    depend on the rank bound $r$.
\end{definition}

The following comparison with the squared trace distance gives the threshold bounds
used in \cref{lem:BQP-hardness_dpower}.

\begin{lemma}\label{lem:dtr_vs_dpower}
    Fix a positive integer $n$ and $p\in\interval[open]{0}{1}$, and let
    $\rho$ and $\sigma$ be $n$-qubit pure density operators. Then
    \begin{equation*}
        \frac{1/p-1}{2^{1/p+1}}
        \dtr{\rho}{\sigma}{2}
        \leq\dpower{\rho}{\sigma}{p}
        \leq\frac{1}{2p}\dtr{\rho}{\sigma}{2}.
    \end{equation*}
\end{lemma}

\begin{proof}
    Write $\rho=\ketbra{\phi}{\phi}$ and
    $\sigma=\ketbra{\psi}{\psi}$ for normalized pure states, and set
    $t\coloneqq\frac{1-\abs{\braket{\phi}{\psi}}}{2}
    \in\interval{0}{1/2}$.
    Since $\rho^p=\rho$ and $\sigma^p=\sigma$, and the only possibly
    nonzero eigenvalues of $\rbra{\rho+\sigma}/2$ are $1-t$ and $t$,
    $\dpower{\rho}{\sigma}{p} =1-\rbra{1-t}^{1/p}-t^{1/p}$.
    The pure-state trace-distance identity gives
    $\dtr{\rho}{\sigma}{2}=1-\abs{\braket{\phi}{\psi}}^2
    =4t\rbra{1-t}$.
    Both quantities vanish at $t=0$; assume $0<t\leq1/2$.

    By Bernoulli's inequality (\cref{fact:Bernoulli_inequality}),
    $\rbra{1-t}^{1/p}\geq1-t/p$, and hence
    $\dpower{\rho}{\sigma}{p}\leq t/p$.
    Applying the mean value theorem (\cref{fact:mean_value_theorem}) to
    $x\mapsto x^{1/p}$ on $\interval{1-t}{1}$ gives
    \begin{equation*}
        1-\rbra{1-t}^{1/p}
        \geq\frac{1}{p}\rbra{1-t}^{1/p-1}t
        \geq\frac{1}{p}2^{1-1/p}t.
    \end{equation*}
    Subtracting $t^{1/p}\leq2^{1-1/p}t$ gives
    $\dpower{\rho}{\sigma}{p}
    \geq\rbra{\frac{1}{p}-1}2^{1-1/p}t$.
    Combining these estimates with
    $2t\leq\dtr{\rho}{\sigma}{2}\leq4t$ proves the claim.
\end{proof}

\begin{lemma}[$\BQP$-hardness for pure states]
\label{lem:BQP-hardness_dpower}
    Fix a polynomial-time computable constant
    $p\in\interval[open]{0}{1}$ and efficiently computable functions
    $a,b$ from the positive integers to $\interval{0}{1}$
    such that $b\rbra{n}<a\rbra{n}$ for every positive integer $n$.
    Suppose that, for all sufficiently large $n$,
    \begin{equation*}
        a\rbra{n}\leq\frac{1/p-1}{2^{1/p+1}}
        \rbra*{1-2^{-n-1}}^2,\qquad
        b\rbra{n}\geq\frac{1}{2p}2^{-2n-2}.
    \end{equation*}
    Then $\PowerPureQSD_p\sbra{a,b}$ is $\BQP$-hard.
\end{lemma}

\begin{proof}
    We reduce from the $\BQP$-hard problem in \cref{lem:BQP-hard}.
    Fix an integer $n_0\geq2$ such that both threshold bounds hold for every
    $n\geq n_0$. For a source instance with $n\geq n_0$, use the identity
    map on the circuit descriptions. Let $\rho$ and $\sigma$ be the pure
    reduced states on the designated output qubits. By
    \cref{lem:dtr_vs_dpower}, the source Yes and No conditions imply,
    respectively,
    \begin{equation*}
        \dpower{\rho}{\sigma}{p}
        \geq \frac{1/p-1}{2^{1/p+1}} \rbra*{1-2^{-n-1}}^2
        \geq a\rbra{n},\qquad
        \dpower{\rho}{\sigma}{p}
        \leq \frac{1}{2p}2^{-2n-2}
        \leq b\rbra{n}.
    \end{equation*}
    For $n<n_0$, the source circuits have bounded width, so classical
    simulation estimates their output trace distance to error $1/8$ in
    polynomial time. Comparing with $1/2$ distinguishes the promised
    cases, whose thresholds are at least $3/4$ and at most $1/4$.
    Output fixed circuits preparing orthogonal $n_0$-qubit computational-basis
    states in the Yes case and identical states in the No case.
    These satisfy the target promises by \cref{lem:dtr_vs_dpower} and
    the threshold bounds at $n_0$. Both $n_0$ and the output circuits
    depend only on $p,a,b$ and can be hard-coded.
    This gives a polynomial-time many-one reduction on all promised
    instances.
\end{proof}

\subsection{\texorpdfstring{$\BQP$}{BQP}-containment}

We next prove containment in $\BQP$ for inverse-polynomial promise gaps,
using the divergence estimator in \cref{thm:query_estimate_power_divergence}.

\begin{lemma}[$\BQP$-containment]
\label{lem:BQP-containment_dpower}
    Fix a polynomial-time computable constant
    $p\in\interval[open]{0}{1}$. Let $r,a,b$ be efficiently computable
    functions on the positive integers, where $r$ is a known positive
    integer-valued rank bound and the threshold functions $a,b$ take
    values in $\interval{0}{1}$. Assume
    $1\leq r\rbra{n}\leq\poly\rbra{n}$ and
    $a\rbra{n}-b\rbra{n}\geq1/\poly\rbra{n}$ for every
    positive integer $n$. Then
    $\PowerLowQSD_p\sbra{a,b}$ is in $\BQP$.
\end{lemma}

\begin{proof}
    The input circuits prepare purifications of $\rho$ and $\sigma$.
    Permuting their output registers gives purified query oracles whose
    direct, inverse, and controlled implementations have polynomial size.

    Fix a positive integer-valued polynomial $g$ satisfying
    $a\rbra{n}-b\rbra{n}\geq1/g\rbra{n}$ for every $n\geq1$, and set
    $\eps\coloneqq1/\rbra{8g\rbra{n}}$. Using the efficient computability
    of $a$ and $b$, compute a rational number $\widetilde m_n$ within
    additive error $\eps$ of $\rbra{a\rbra{n}+b\rbra{n}}/2$ in polynomial
    time. Run the estimator in \cref{thm:query_estimate_power_divergence} with
    rank bound $r\rbra{n}$ and accuracy $\eps$, and let $x$ be its output.
    With probability at least $2/3$,
    $\abs{x-\dpower{\rho}{\sigma}{p}}\leq\eps$.

    Condition on this event. On a Yes instance, $x-\widetilde m_n$ is at
    least $\frac{a\rbra{n}-b\rbra{n}}{2}-2\eps\geq2\eps$.
    The same lower bound holds for $\widetilde m_n-x$ on a No instance.
    Thus the rule that returns Yes exactly when $x\geq\widetilde m_n$
    succeeds with probability at least $2/3$.

    Since $r\rbra{n},1/\eps\leq\poly\rbra{n}$ and $p$ is fixed,
    \cref{thm:query_estimate_power_divergence} gives polynomial query
    complexity. Its polynomial construction and precision choices, together
    with \cref{thm:qsvt,lem:eigen_threshold_projector,lem:trace_est_subdensity-operator},
    give polynomial non-oracle gate counts and classical circuit-generation
    time. The state-preparation, LCU, evolution, and input-oracle circuits
    also have polynomial overhead, yielding a uniform polynomial-time
    quantum algorithm.
\end{proof}

\begin{proof}[Proof of \cref{thm:completeness_lowrank_dpower}]
    The pure-state restriction is $\BQP$-hard by
    \cref{lem:BQP-hardness_dpower}. Since $r\rbra{n}\geq1$, the identity
    map on its circuit descriptions is a polynomial-time many-one
    reduction to the low-rank problem, so the low-rank problem is
    $\BQP$-hard.

    Since $a\rbra{n}>b\rbra{n}$ for every $n$, enlarging the polynomial
    in the eventual inverse-polynomial gap bound gives such a bound for all
    $n\geq1$. Containment in $\BQP$ now follows from
    \cref{lem:BQP-containment_dpower}.
\end{proof}

\subsection*{Acknowledgments}
J.~B. is supported by the Quantum Advantage Turbo Charger (QATCH)
programme, funded by UKRI under Grant No.~UKRI4257, and by the
Quantum Advantage Pathfinder (QAP) programme, funded by EPSRC under
Grant No.~EP/X026167/1.

\subsection*{AI Use Statement}
During the preparation of this manuscript, the authors used OpenAI Codex
to assist with language editing, notation consistency checks, LaTeX
debugging, and the review and revision of mathematical arguments.
The authors reviewed the resulting suggestions and take full
responsibility for the final content of the manuscript.

\phantomsection
\addcontentsline{toc}{section}{References}

\bibliographystyle{alphaurl}
\bibliography{references}

\end{document}